\documentclass[journal]{vgtc}                

\newcommand{\final}{1}

\ifpdf
  \pdfoutput=1\relax                   
  \usepackage{graphicx}                
  \DeclareGraphicsExtensions{.pdf,.png,.jpg,.jpeg} 
\else
  \usepackage{graphicx}                
  \DeclareGraphicsExtensions{.eps}     
\fi%

\graphicspath{{figures/}{pictures/}{images/}{./}{../}} 

\usepackage{microtype}                 
\PassOptionsToPackage{warn}{textcomp}  
\usepackage{textcomp}                  
\usepackage{mathptmx}                  
\usepackage{times}                     
\usepackage{cite}                      
\usepackage{tabu}                      
\usepackage{booktabs}                  

\usepackage{color}
\usepackage{ifthen}
\usepackage{float}
\usepackage{alltt}
\usepackage{amsmath}
\usepackage{amssymb}
\usepackage{amsthm}
\usepackage{newlfont} 
\usepackage{floatflt}
\usepackage{wrapfig}
\usepackage{fixltx2e}
\usepackage{subfig} 
\usepackage{multirow}
\usepackage{booktabs}
\usepackage{cleveref}
\usepackage{algorithmic}
\usepackage[export]{adjustbox}
\usepackage{mathtools, cuted}
\usepackage{CJKutf8} 
\usepackage{epsfig}
\usepackage{soul}
\usepackage{afterpage}

\usepackage[ruled]{algorithm2e} 

\usepackage{array} 

\SetAlFnt{\small}
\SetAlCapFnt{\small}
\SetAlCapNameFnt{\small}
\SetAlCapHSkip{0pt}
\IncMargin{-\parindent}
\newcommand{\Caption}[2]{\caption[#1]{{\em #1} #2}}
\let\oldcaption\caption
\renewcommand{\caption}[2][]{\oldcaption[#1]{{\em #1} #2}}

\definecolor{SithColor}{rgb}{0.7,0,0} 
\newcommand{\qisun}[1]{{\color{SithColor} Qi: #1 $\qed$}}
\newcommand{\monde}[1]{{\color{SithColor} Monde: #1 $\qed$}}
\definecolor{nyuColor}{rgb}{0.34,0.18,0.55} 
\newcommand{\zh}[1]{{\color{nyuColor} Zhenyi: #1 $\qed$}}
\definecolor{sjtuColor}{rgb}{0.31,0.65,0}
\newcommand{\dnc}[1]{{\color{sjtuColor} Nianchen: #1 $\qed$}}

\newcommand{\zhtd}[1]{{\color{nyuColor} #1 }}
\definecolor{ConsularColor}{rgb}{0,0.4,0} 
\definecolor{GuardianColor}{rgb}{0,0,0.8} 
\newcommand{\praneeth}[1]{{\color{GuardianColor} Praneeth: #1 $\qed$}}
\newcommand{\new}[1]{{\color{ConsularColor}#1 $\qed$}}
\newcommand{\rev}[1]{{#1}}
\newcommand{\warning}[1]{{\it\color{red} #1}}
\newcommand{\note}[1]{{\it\color{blue} #1}}
\newcommand{\nothing}[1]{}

\definecolor{AudioColor}{rgb}{0.56,0.34,0.62}

\definecolor{figred}{rgb}{1,0,0}
\definecolor{figgreen}{rgb}{0,0.6,0}
\definecolor{figblue}{rgb}{0,0,1}
\definecolor{figpink}{rgb}{1,0.63,0.63}

\ifthenelse{\equal{\final}{1}}
{
\renewcommand{\qisun}[1]{}
\renewcommand{\zh}[1]{}
\renewcommand{\dnc}[1]{}
\renewcommand{\zhtd}[1]{}
\renewcommand{\praneeth}[1]{}
\renewcommand{\new}[1]{{#1}}
\renewcommand{\warning}[1]{}
\renewcommand{\note}[1]{}
\renewcommand{\monde}[1]{}
}
{}

\newcommand{\pseudocode}{Algorithm}
\floatstyle{plain}
\newfloat{algorithm}{tbhp}{lop}
\floatname{algorithm}{\pseudocode}

\newcommand{\filename}[1]{\url{#1}}
\newcommand{\foldername}[1]{\url{#1}}

\let\oldparagraph\paragraph
\iffalse

\renewcommand{\paragraph}[1]{\oldparagraph{\textbf{#1}.}} 
\else

\renewcommand{\paragraph}[1]{\oldparagraph{{#1}.}}
\fi

\ifdefined\email
\else
\newcommand{\email}[1]{\url{#1}}
\fi

\newcommand{\sparseError}{E_{scene}}
\newcommand{\pt}{\mathbf{q}}
\newcommand{\rayo}{\mathbf{x}}
\newcommand{\rayd}{\mathbf{v}}
\newcommand{\intersectionFunc}{p}

\newcommand{\SpatialPt}{\mathbf{q}}

\newcommand{\sphereNum}{N}

\newcommand{\sphereRadius}{r}

\newcommand{\projectionMatrix}{\mathbf{M}}
\newcommand{\imgSpaceError}{E_{image}}

\newcommand{\finalError}{E}
\newcommand{\latency}{l}

\newcommand{\mlpLayerNum}{N_{m}}
\newcommand{\mlpChannelNum}{N_{c}}

\newcommand{\imageFoveal}{I_{f}}
\newcommand{\imageMid}{I_{m}}
\newcommand{\imageFar}{I_{p}}

\newcommand{\mlpFunc}{\mathcal{C}}

\newcommand{\norm}[1]{\left\lVert#1\right\rVert}

\newcommand{\argmin}{\operatornamewithlimits{argmin}}

\newcommand{\camDir}{\mathbf{R}}
\newcommand{\gazeDir}{\mathbf{\rayd_g}}

\onlineid{1024}

\vgtccategory{Research}
\vgtcpapertype{algorithm/technique}

\title{FoV-NeRF: Foveated Neural Radiance Fields for Virtual Reality} 

\author{Nianchen Deng$^*$, Zhenyi He$^*$, Jiannan Ye, Budmonde Duinkharjav, Praneeth Chakravarthula, Xubo Yang$^\dagger$, and Qi Sun$^\dagger$}
\authorfooter{
\item Nianchen Deng and Jiannan Ye are with School of Software, Shanghai Jiao Tong University. E-mail: \{dengnianchen, wsyhdyjn\}@sjtu.edu.cn.
\item Zhenyi He is with Department of Computer Science, New York University. E-mail: zh719@nyu.edu.
\item Budmonde Duinkharjav is with Immersive Computing Lab, New York University. E-mail: budmonde@gmail.com.
\item Praneeth Chakravarthula is with Computer Science, UNC Chapel Hill. E-mail: cpk@cs.unc.edu
\item Xubo Yang is with School of Software, Shanghai Jiao Tong University and Peng Cheng Laboratory. E-mail:  yangxubo@sjtu.edu.cn.
\item Qi Sun is with Tandon School of Engineering, New York University. E-mail: qisun@nyu.edu.
\item[*] These authors contributed equally to this work.
\item[$\dagger$] The corresponding authors.
}

\shortauthortitle{Deng \MakeLowercase{\textit{et al.}}: Fov-NeRF: Foveated Neural Radiance Field for Virtual Reality}

\abstract{
Virtual Reality (VR) is becoming ubiquitous with the rise of consumer displays and commercial VR platforms. Such displays require low latency and high quality rendering of synthetic imagery with reduced compute overheads. 
\new{Recent advances in neural rendering showed promise of unlocking new possibilities in 3D computer graphics via image-based representations of virtual or physical environments.
Specifically, the neural radiance fields (NeRF) demonstrated that photo-realistic quality and continuous view changes of 3D scenes can be achieved without loss of view-dependent effects. 
While NeRF can significantly benefit rendering for VR applications, it faces unique challenges posed by high field-of-view, high resolution, and stereoscopic/egocentric viewing, typically causing low quality and high latency of the rendered images. In VR, this not only harms the interaction experience but may also cause sickness.

To tackle these problems toward six-degrees-of-freedom, egocentric, and stereo NeRF in VR, we present \emph{the first gaze-contingent 3D neural representation and view synthesis method}. We incorporate the human psychophysics of visual- and stereo-acuity into an egocentric neural representation of 3D scenery. We then jointly optimize the latency/performance and visual quality while mutually bridging human perception and neural scene synthesis to achieve perceptually high-quality immersive interaction. 
We conducted both objective analysis and subjective studies to evaluate the effectiveness of our approach. We find that our method significantly reduces latency (up to 99\% time reduction compared with NeRF) without loss of high-fidelity rendering (perceptually identical to full-resolution ground truth). 
The presented approach may serve as the first step toward future VR/AR systems that capture, teleport, and visualize remote environments in real-time.}

}

\keywords{Virtual Reality; Gaze-Contingent Graphics; Neural Representation; Foveated Rendering}

\teaser{
  \centering
  \subfloat[scene representation]{
    \label{fig:teaser:scene}
    \includegraphics[height=3.7cm]{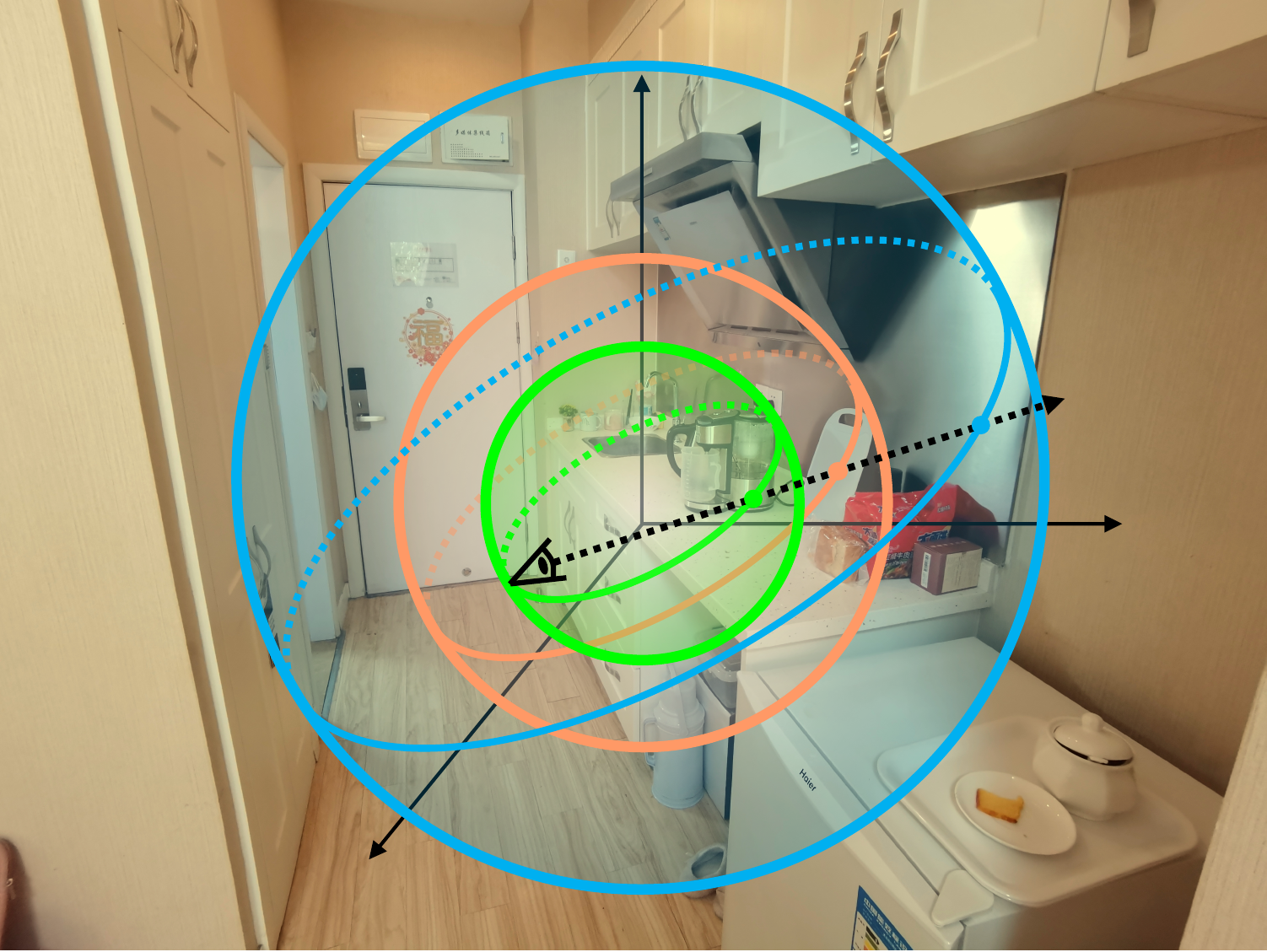}
  }
  \subfloat[overall quality]{
    \label{fig:teaser:latency}
    \includegraphics[height=3.7cm]{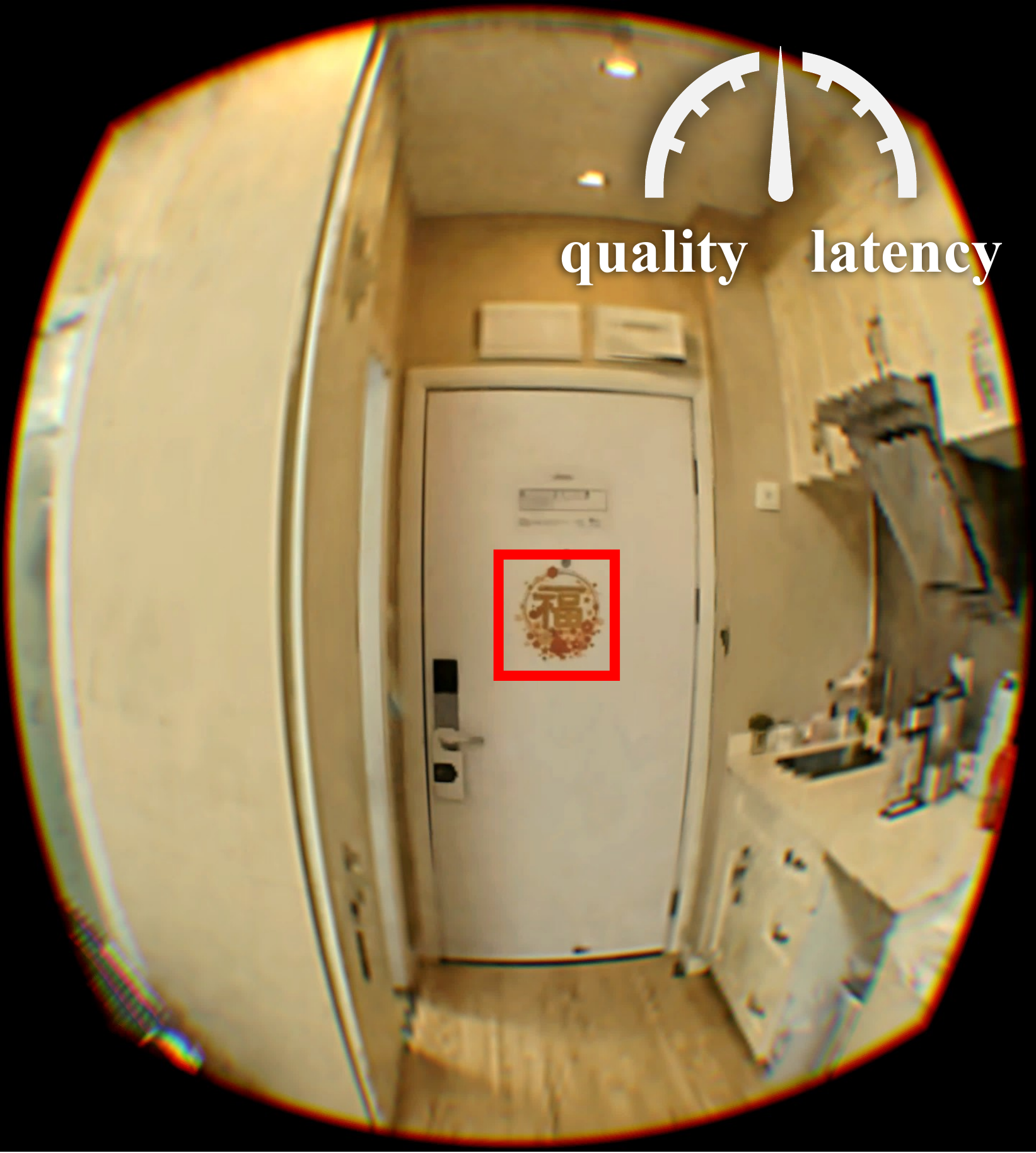}
  }
  \subfloat[foveal quality]{
    \label{fig:teaser:quality}
    \includegraphics[height=3.7cm]{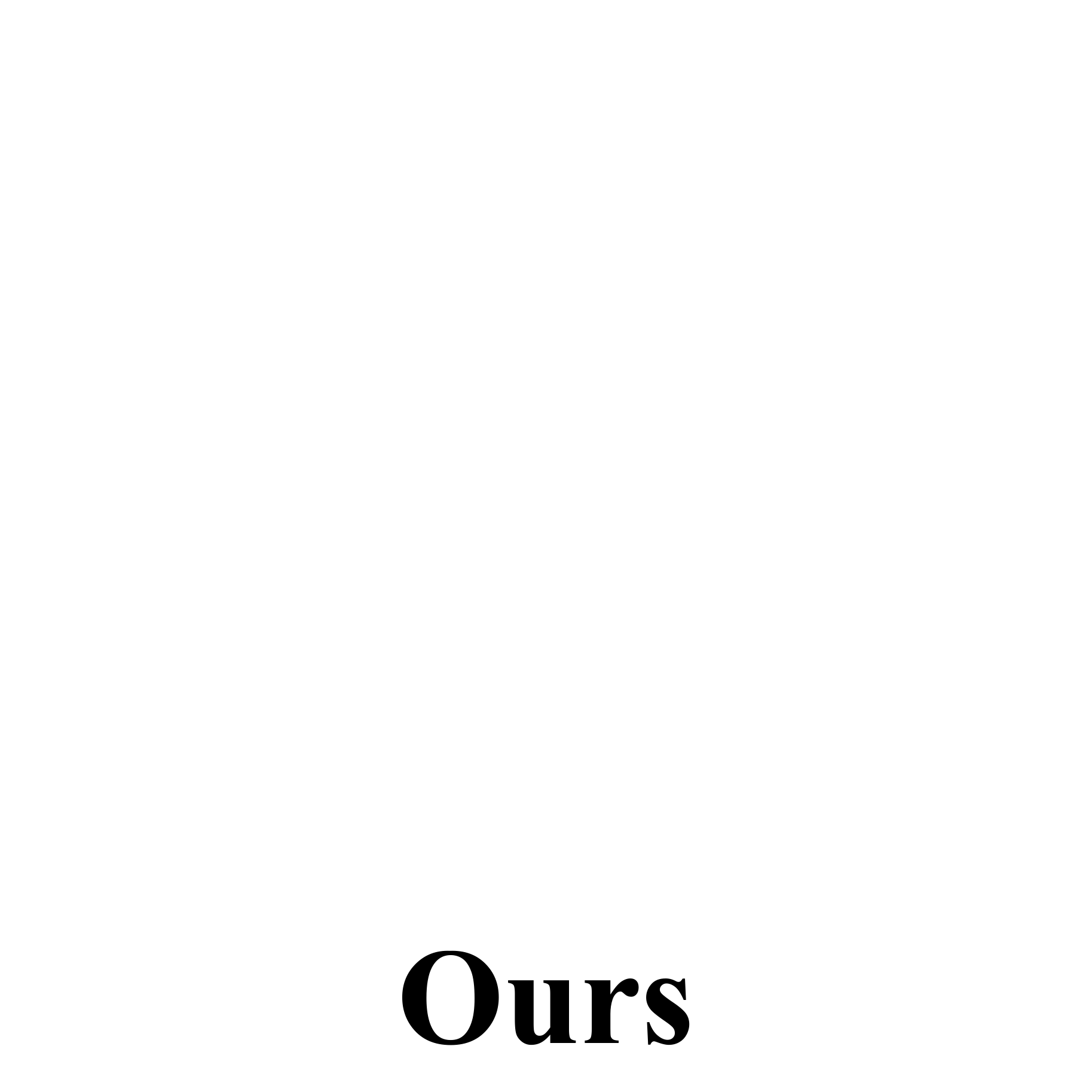}
  }
   \subfloat[foveal references]{
    \label{fig:teaser:reference}
    \includegraphics[height=3.7cm]{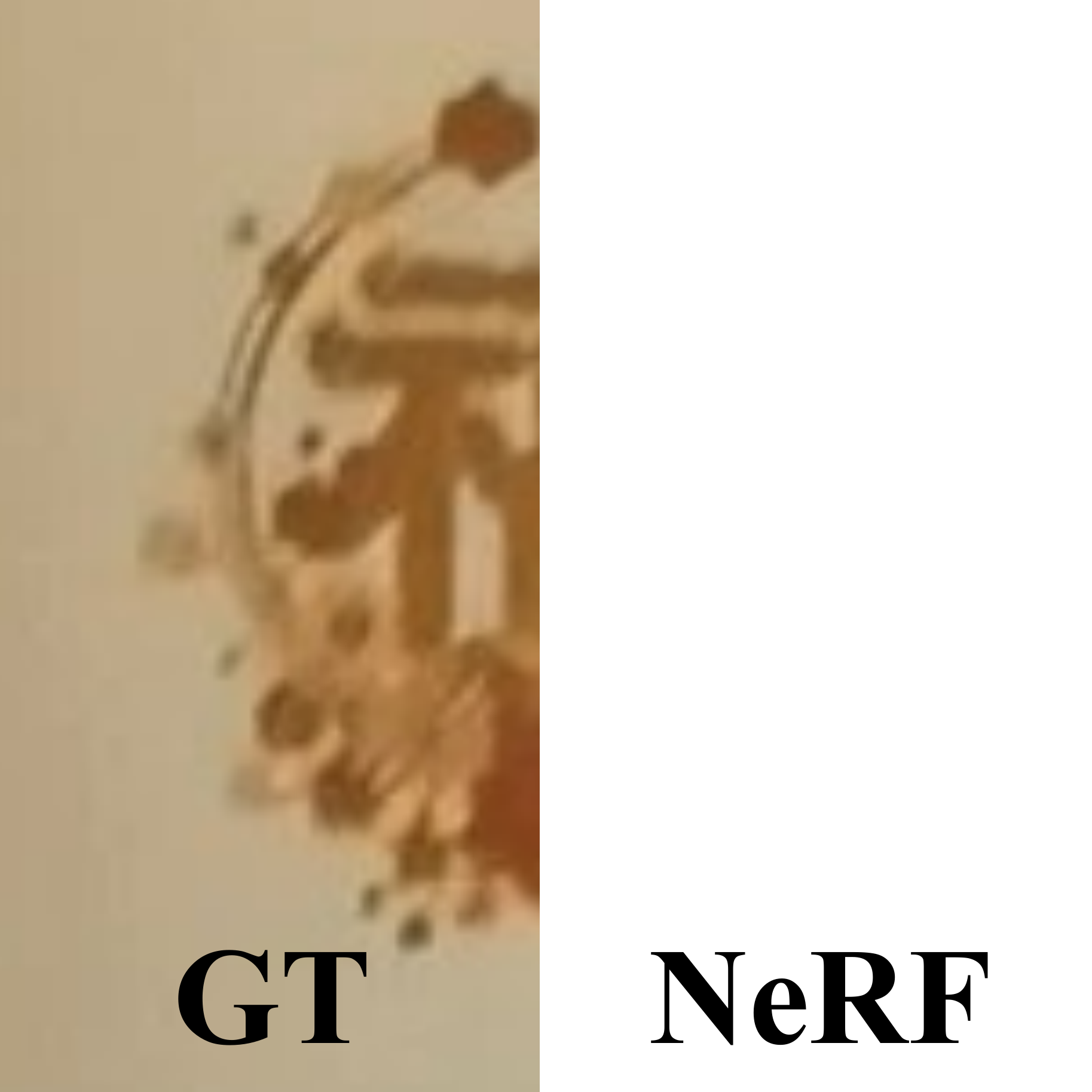}
  }
 \Caption{Illustration of our gaze-contingent neural radiance field for VR.}
 {%
\subref{fig:teaser:scene} Visualization of our egocentric-viewing-tailored coordinate and neural scene representation.
\subref{fig:teaser:latency} 
Our neural encoding and synthesis method matches human visual and stereoscopic acuity and also balances the quality and run-time perceptual latency.
Beyond fast performance, \subref{fig:teaser:quality} and \subref{fig:teaser:reference} zoom into the foveal region where the human vision has highest sensitivity to imagery quality and stereopsis. 
\textbf{Our} method shows superior quality than alternative neural rendering (\textbf{NeRF} as in \subref{fig:teaser:reference}, \protect\cite{mildenhall2020nerf}) approaches, with the full quality rendered ground truth, \textbf{GT}, as reference \subref{fig:teaser:reference}.
 }
 \label{fig:teaser}
}

\begin{document}

\firstsection{Introduction}
\maketitle
\note{
What problem we are trying to solve.
Why it is important, and why people should care.
}

\new{
The unprecedented virtual and augmented reality (VR and AR) hardware revolutions have enabled realistic immersion and natural interaction. An essential need for future VR/AR is to teleport different users and different environments for collaborative scenarios. The recent advancement of neural radiance field (NeRF) methods encode a full radiance field reconstruction via image-based shots \cite{mildenhall2020nerf}. The high quality and flexibility show promising applications for serving as future VR content creation and consumption representations. 

Unlike traditional screens, practical VR displays require high field-of-view (FoV), low latency, stereoscopic, and egocentric viewing. These requirements are fundamental to ensure realism and safety without strong simulator sickness. However, the inference from NeRF representation, compared with forward-rendering, demands heavy computation and causes high latency and/or low quality while consumed with immersive viewers.
}

\note{
What prior works have done, and why they are not adequate.
(Note: this is just high level big ideas. Details should go to a previous work section.)
}

Acceleration without compromising perceptual quality is the ultimate goal of real-time rendering.
Gaze-contingent rendering approaches have shown remarkable effectiveness in presenting imagery of perceptual quality identical to full resolution rendering~\cite{Guenter:2012:F3G,Patney:2016:TFR,Tursun:2019:LCA}. This is achieved by harnessing high FoV head-mounted displays and high precision eye-tracking technologies. \new{However, existing forward foveated rendering approaches rely on 3D assets such as triangularized geometries. Acquiring such information for physical world objects, such as humans, typically contain noise or low-quality texture. This problem, in comparison, can be harnessed by image-based approaches.}

There has been a surge in image-based neural representations/rendering as an alternative to traditional 3D representations such as polygonal meshes. Examples include voxelization \cite{sitzmann2019deepvoxels}, distance fields \cite{sitzmann2019metasdf,mildenhall2020nerf,park2019deepsdf}, multi-layer panorama imagery \cite{Lin:DeepPanorama,Attal:2020:ECCV}, and many more. 
However, existing solutions suffer from high time consumption (i.e., latency), low image fidelity, or inconsistent robustness of complex scenes with occlusion. 
\new{Recent accelerations include depicting the inference as basis functions \cite{Wizadwongsa2021NeX} or fast integration via learning gradients \cite{autoint}. Whereas, no method encodes the human visual system and optimizes for the egocentric and stereoscopic viewing in VR.}

\note{
What our method has to offer, sales pitch for concrete benefits, not technical details.
Imagine we are doing a TV advertisement here.
}


\new{
We present the first gaze-contingent neural radiance representation and foveated synthesis approach. 
By encoding human vision's spatial/stereoscopic acuity and temporal sensitivity, we significantly improve responsiveness (from 9s to 20 ms) from alternative neural synthesis methods. This is achieved without loss of visual fidelity compared with the rendered or captured ground truth.
}

\note{
Our main idea, giving people a take home message and (if possible) see how clever we are.
}
\new{
We achieve this by first representing 3D radiance fields with concentric spherical coordinates. The tailored representation both optimizes for egocentric viewing and minimizes inference running time.
Our representation also allows for depicting the foveated color and stereopsis sensitivities during the training and inference phases. 
To this end, we devise retina-matched perceptual models and VR-tailored representation coordinates toward high visual quality as well as low systematic latency.
Lastly, we derive an analytical spatial-temporal perception model from optimizing our neural scene representation toward imperceptible loss in image quality and latency.
}

\note{
Our algorithms and methods to show technical contributions and that our solutions are not trivial.
}



\note{
Results, applications, and extra benefits.
}
We validate our system by conducting psychophysical experiments, numerical analysis, and case studies on commercially available VR display devices. 
The series of experiments reveals our method's effectiveness and advantages: it delivers immersive, high-FoV, and high-fidelity neural radiance fields that are perceived identical to an arbitrarily high-quality rendered 3D scene with perception optimized quality-latency performance. To encourage third-party reproduction and extensions in the community, we will open-source our implementation and dataset. \rev{The code and dataset are accessible at https://github.com/dengnianchen/fovnerf.}
In summary, we make the following major contributions:
\new{
\begin{itemize}
\itemsep -0.5\parsep
    \item A low-latency and high-fidelity immersive application, offering full perceptual quality with instant high resolution and high FoV first-person VR viewing;
    \item A 3D neural radiance field representation tailored for egocentric immersive viewing and accelerated inference;
    \item A human-vision-matched neural synthesis method considering both visual- and stereo- acuity;
    \item A spatio-temporal analytical model for jointly optimizing systematic latency and perceptual quality for human observers.
\end{itemize}
}

\section{Related Work}
\label{sec:prior}
\subsection{Image-based View Synthesis}
Image-based rendering (IBR) has been proposed in computer graphics to complement the traditional 3D assets and ray propagation pipelines \cite{levoy1996light,gortler1996lumigraph,shum2000review}. Benefiting from the flexible 2D representation, it delivers pixel-wise high quality without compromising rendering performance. 
However, a major limitation of IBR is its sparse and small viewing range, narrowing the applicability in the smooth VR viewing experience where the head/gaze continuously moves.
To address this problem, recent research has been exploring synthesizing novel image-based views instead of fully capturing them. Examples include synthesizing light fields \cite{LearningViewSynthesis,Li2020LF,mildenhall2019llff} and multi-layer volumetric videos \cite{Broxton:immersiveLF}. 
However, the synthesis usually suffers from the trade-off between optimal quality and fast performance.

\subsection{Neural Scene Representation and Rendering}
To fully represent a 3D object and environment, neural representations have drawn extensive attention. 
With a deep neural network that depicts a 3D world as an implicit function, the neural networks may directly approximate an object's appearance given a camera pose \cite{sitzmann2019deepvoxels,sitzmann2019srns,sitzmann2019siren,mildenhall2020nerf,park2017transformation}. 
Prior arts also investigated implicitly representing shapes ~\cite{park2019deepsdf} and surfaces ~\cite{mescheder2019occupancy} to improve the visual quality of 3D objects. Inspired by \cite{park2019deepsdf}, signed distance functions have been deployed for representing raw data such as point cloud ~\cite{atzmon2020sal,gropp2020implicit} with better efficiency ~\cite{chabra2020deep}. 
The input information ranged from 2D images ~\cite{lin2020sdf, yariv2020multiview,choi2019extreme}, 3D context ~\cite{saito2019pifu, oechsle2019texture, zhang2020deep}, time-varying vector field~\cite{niemeyer2019occupancy} to local features~\cite{tretschk2020patchnets, liu2020neural}.
Current implicit representations primarily focus on locally ``outside-in'' viewing of individual objects, with low field-of-view, speed, and resolution. For large scenes, the lack of coverage may cause the quality to drop.
However, in virtual reality, the cameras are typically first-person and highly dynamic, requiring low latency, high resolution, and 6DoF coverage. 

\subsection{Panorama-Based 6DoF Immersive Viewing}
Panorama images and videos are directly applicable to VR platforms thanks to their 360 FoV coverage. 
However, the main challenge is their single projection center in each frame, limiting free-form camera translation, thus 6DoF natural viewing experience. 
Recently, extensive research has been proposed to address this problem. Serrano et al. \cite{Serrano_TVCG_VR-6dof} presented a depth-based dis-occlusion method that dynamically reprojects to 6DoF cameras, enabling natural viewing with motion parallax. Pozo et al. \cite{Pozo:2019:I6V} jointly optimizes capture and viewing processes.
Machine learning approaches have advanced the robustness of various geometric and lighting conditions \cite{Attal:2020:ECCV,Lin:DeepPanorama,Benjamin:2020:RTV}. 
As a spherical image, panoramas are designed for capturing physical worlds with flexible reprojections to displays. The 6DoF viewing typically suffers from trade-offs among performance, achievable resolution/translation ranges, and dis-occlusion artifacts due to the insufficient none-light-of-sight capture.
With an orthogonal mission, our model represents and reproduces arbitrary 3D virtual or physical worlds. 
In \Cref{sec:study:quality}, we compare our method with a panoramic-imagery-based view synthesis approach considering visual quality and allowable translation.

\subsection{Gaze-Contingent Rendering}
Both rendering and neural synthesis suffer from the heavy computational load, thus systematic latency and lags.
With a full mesh representation, foveated rendering has been proposed to accelerate local rendering performance and/or enhancing perceptual cues, including mesh- \cite{Guenter:2012:F3G,Patney:2016:TFR,Konrad:2019:OcularParallax,chen2022instant}, image- \cite{Kaplanyan:2019:DNR,Krajancich:2020:gc_stereo}, and optically-based \cite{Sun:2017:PGF,chakravarthula2021gaze} methods. 
The accelerations are typically achieved through high-FoV displays (such as VR headsets) and eye-tracking technologies~\cite{lu2020improved}. However, these all require full access to the original 3D assets. 
Our method bridges human vision and neural rendering.
It accelerates and extends neural synthesis beyond local rendering, enabling  computational efficiency and perceptually high fidelity.
\section{Method}\label{sec:method}

\begin{figure}[tb]
    \centering
    \includegraphics[width=\linewidth]{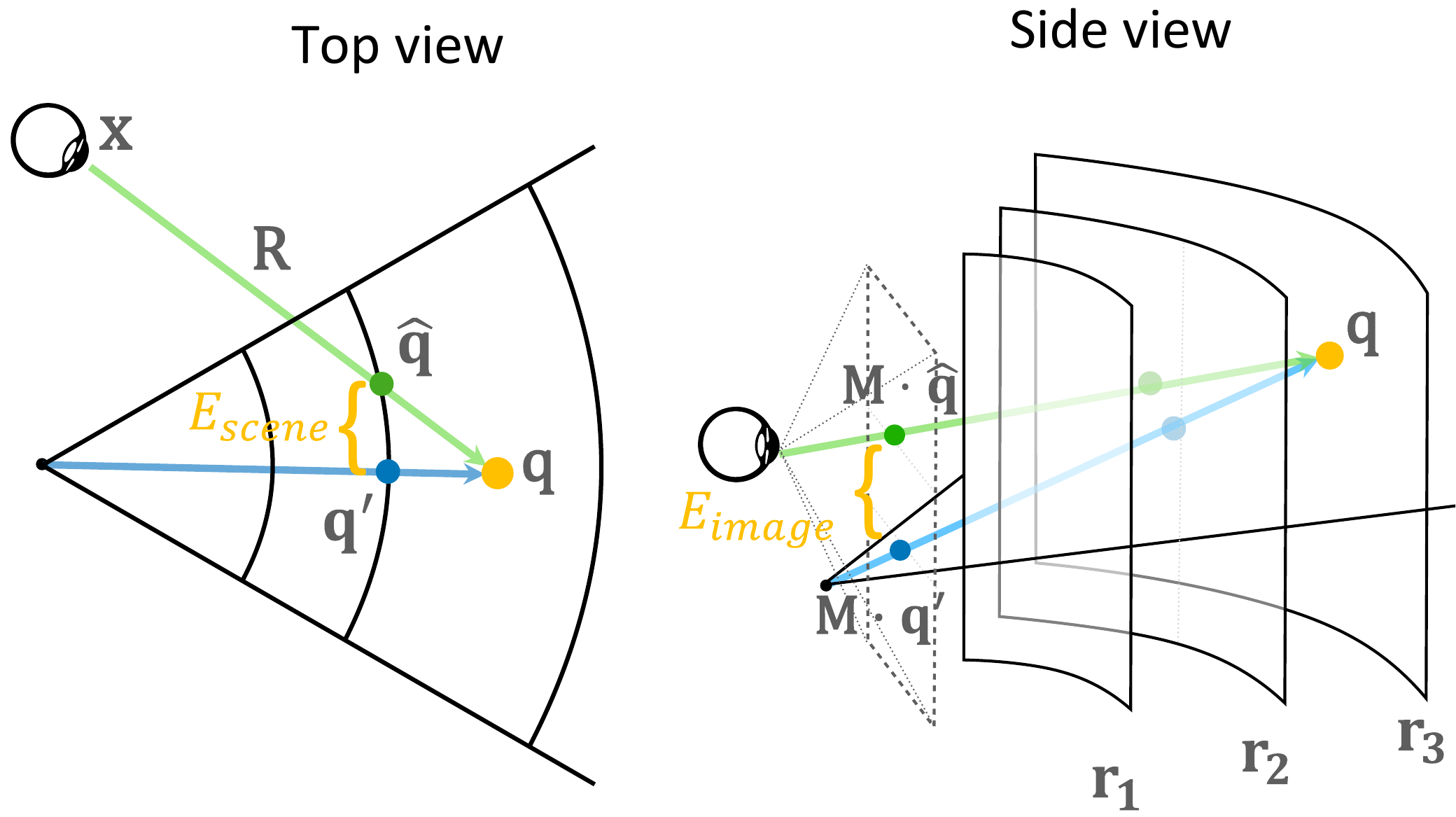}
    
    \Caption{Coordinate system and variable annotations.}
    {%
    We partially visualize our 3D full spherical representation with an example of $\sphereNum=3$ spheres. Variables and equations are annotated at corresponding positions.
    }
    \label{fig:notations}
\end{figure}

Based on a concentric spherical representation and correspondingly trained network predicting RGBDs (\Cref{sec:method:representation}), our system predominantly comprises two main rendering steps in run-time: synthesizing visual-/stereoscopic-acuity-adaptive elemental images with ray marching for fovea, mid-, and far- peripheries; followed by image-based rendering to composite displayed frames (\Cref{sec:synthesis}).
For desired precision-performance balance, we further craft an analytical spatial-temporal model to optimize the determination of our intra-system variables, including representation sparsity and neural-network complexities (\Cref{sec:method:optimization}).

\subsection{Egocentric Neural Representation and Training}\label{sec:method:representation}

The recent single-object-oriented ``outside-in'' view synthesis methods \cite{sitzmann2019deepvoxels,mildenhall2020nerf} typically represent the training targets using uniform voxelization. However, immersive VR environments introduce unique and open challenges for such parameterization due to the commonly egocentric (first-person) and ``inside-out'' viewing perspective (e.g., \Cref{fig:teaser:scene}).
As a consequence, the neural representation on large virtual environments typically suffer from ghosting artifacts, low resolution, or slow speed (\Cref{fig:teaser:quality}).

\paragraph{Egocentric coordinate}
To tackle this problem, we are inspired by the recent panoramic imagery dis-occlusion methods \cite{Lin:DeepPanorama,Benjamin:2020:RTV,Broxton:immersiveLF}: we depict the rapidly varying first-person views with concentric spherical coordinates. This representation has been shown to allow for robust rendering at real-time rates and 6DoF interaction to navigate inside complex immersive environments. 
As visualized in \Cref{fig:notations}, our representation is parameterized with the number of concentric spheres per neural network ($\sphereNum$) and their respective radii ($\mathbf{\sphereRadius}=\{\sphereRadius_i\},i\in [1,\sphereNum]$). \new{Under this spherical system, a given 3D spatial position can be represented as $\SpatialPt=\SpatialPt(\sphereRadius,\theta,\phi)$ where $\theta$ and $\phi$ are two angular numbers.
Similar to \cite{mildenhall2020nerf}, the run-time rendering goal is predicting a 4D vector $(r,g,b,d)$ for each $\SpatialPt$, followed by a view-dependent ray marching through this intermediate function to synthesize individual pixel.}
Here, $(r,g,b)$ and $d$ are the color and density, respectively. 

\qisun{(May 15) double check and revise this paragraph, esp the beginning claims. I may want to add a small figure inset to explain the reasonale.}

\new{
\paragraph{Neural representation}
Existing neural rendering for ``outside-in'' viewing independently train neural networks to predict the 4D vector for individual $\SpatialPt$. This is due to the high variations in view points other than the viewing targets. 
However, egocentric viewing is the opposite: despite the translation for 6DoF viewing, the observers' may change viewing targets frequently by rotating the head and gaze (See \Cref{fig:notations}).
That is, given a neural network's capability, a local egocentric neural representation may encode less viewing changes but more spatial variances. 
To achieve this aim, we design the network to infer an array of vectors per viewing ray:

\begin{equation}
    \mlpFunc(\rayo, \rayd) \triangleq (\mathcal{R},\mathcal{G},\mathcal{B},\mathcal{D}).
    \label{eq:vector_rep}
\end{equation}

Here, $\rayo/\rayd$ defines a ray's origin/direction. $\mathcal{R},\mathcal{G},\mathcal{B},\mathcal{D}$ are $\sphereNum$ dimensional vectors representing the R/G/B/intensities of the $\sphereNum$ intersecting points ($\SpatialPt({\sphereRadius_i,\theta_i,\phi_i}), i\in[1,\sphereNum]$) between the ray and the concentric spheres. The system can then render individual color channels via integrating over $\mathcal{R}/\mathcal{G}/\mathcal{B}$ with $\mathcal{D}$ as weights.
Denser spherical sampling (i.e., higher $\sphereNum$) lead to more precise quality but slower run-time performance. 

Inspired by NeRF \cite{mildenhall2020nerf}, we devise a machine learning approach that encodes $\mlpFunc$ as a multi-layer perception (MLP) neural network. 
NeRF predicts each individual intersection point $\{\mathbf{\hat{\SpatialPt}}_i\}$ before calculates the final color, causing slow performance due to the $\sphereNum$ network inference operation per ray in run-time. 
Our egocentric-viewing-tailored representation (\Cref{eq:vector_rep}) concatenate all $\sphereNum$ encoded coordinates to a vector and feed the coordinates to the MLP module with only one inference.
During training, we define the input $\{\rayo, \rayd\}$ as the $\sphereNum$ intersecting points $\mathbf{\hat{\SpatialPt}_i}$. Our MLP module contains $\mlpLayerNum$ fully-connected layers with $\mlpChannelNum$ channels in each layer. 
}


\subsection{Gaze-Contingent Synthesis during Run-time}\label{sec:synthesis}

Our concentric spherical coordinate, as described in \Cref{sec:method:representation}, addresses the large view target variance problem in the training stage. 
However, it may still suffer from significant rendering latency (about half a second for each stereo frame). This is another essential challenges causing neural representation to be unsuitable yet for immersive viewing. 
In this research, we leverage the spatially adaptive human visual- and stereoscopic- sensitivities to unlock fast runtime inference. 
Instead of the typical single image prediction, we synthesize multiple elemental images to enable real-time responsiveness. The elemental images are generated based on the viewer's head and gaze motions and are adapted to the retinal acuity in resolution and stereo. 
\subsubsection{Adaptive Monoscopic Acuity}

\begin{figure*}[htb]
    \centering
    \includegraphics[width=\linewidth]{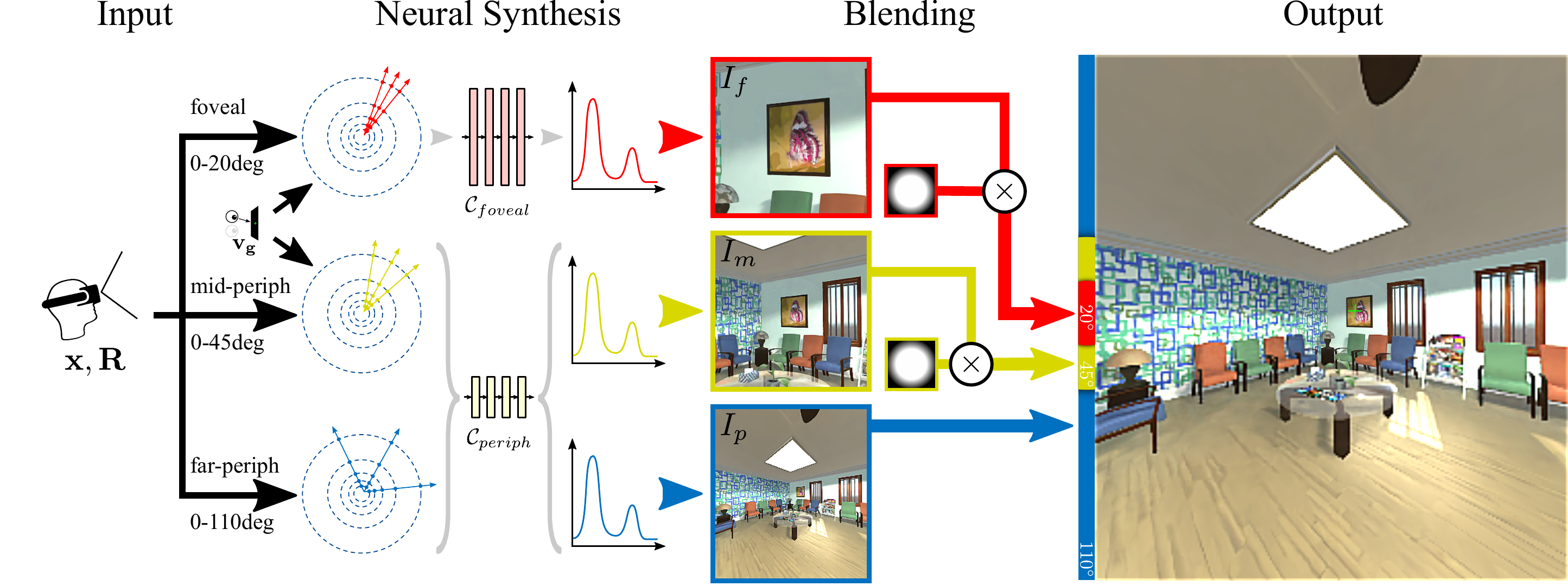}
    \Caption{\rev{Visual acuity adaptive synthesis and rendering mechanism.}}
    {%
    \rev{We first synthesize the elemental images from our egocentric neural representation for fovea (within $20$ deg)/mid-periphery (within $45$ deg)/far-periphery (within $110$ deg, the capability of our VR display). Then these images are blended to the final displayed frame.}
    \nothing{(a)/(b)/(c) shows our individual gaze-contingent synthesis for fovea (within $20$ deg)/mid-periphery (within $45$ deg)/far-periphery (within $110$ deg, the capability of our VR display), respectively. (d) illustrates our real-time rendering system by dynamically blending the individual images to the final displayed frame.}
    }
    \label{fig:system}
\end{figure*}

The human retinal ganglion cells, which collect and transmit visual information, are not uniformly distributed. Instead, its density in the visual fields close to the retinal center is much higher than the periphery \cite{watson2014formula}. This is referred as foveated vision in accelerating immersive rendering \cite{Patney:2016:TFR,Kaplanyan:2019:DNR}.
Inspired by this, we significantly accelerate the runtime inference by integrating the characteristic of spatially-adaptive visual acuity without compromising the {\it perceptual} quality. 

\new{Specifically, given device-tracked camera position ($\rayo$), direction ($\camDir$), and gaze position ($\gazeDir$), we synthesize three elemental images cover different eccentricity ranges of field-of-view (FoV) for each eye: the fovea ($\imageFoveal(\rayo,\camDir,\gazeDir)$, $0-20$ deg), the mid-eccentricity ($\imageMid(\rayo,\camDir,\gazeDir)$, $0-45$ deg) and the entire visual field ($\imageFar(\rayo,\camDir$), $0-110$ deg) (\Cref{fig:system}). Note that $\imageFar$ is independent from the gaze direction $\gazeDir$.
Given the decreasing visual acuity from low to high eccentricities, we devise two orthogonal networks with different hyper parameters: $\mlpFunc_{foveal}$ for $\imageFoveal$ (the high acuity fovea), and $\mlpFunc_{periph}$ for $\imageMid$ and $\imageFar$ (reduced acuity).}

To incorporate the display capabilities and aspect ratios, we define the resolutions of $\imageFoveal$/$\imageMid$/$\imageFar$ as \new{$256^2$/$256^2$/$230\times256$}.
That is, the $\imageFoveal$ has the highest \rev{angular} resolution of \new{$12.8$} pixels per degree (PPD), higher than those of $\imageMid$ ($5.7$ PPD) and $\imageFar$ ($2.33$ PPD).

\subsubsection{Adaptive Stereoscopic Acuity}

\begin{figure}[tb]
    \centering

    \subfloat[\rev{w/o adaptive stereo}]{
        \includegraphics[width=0.48\linewidth]{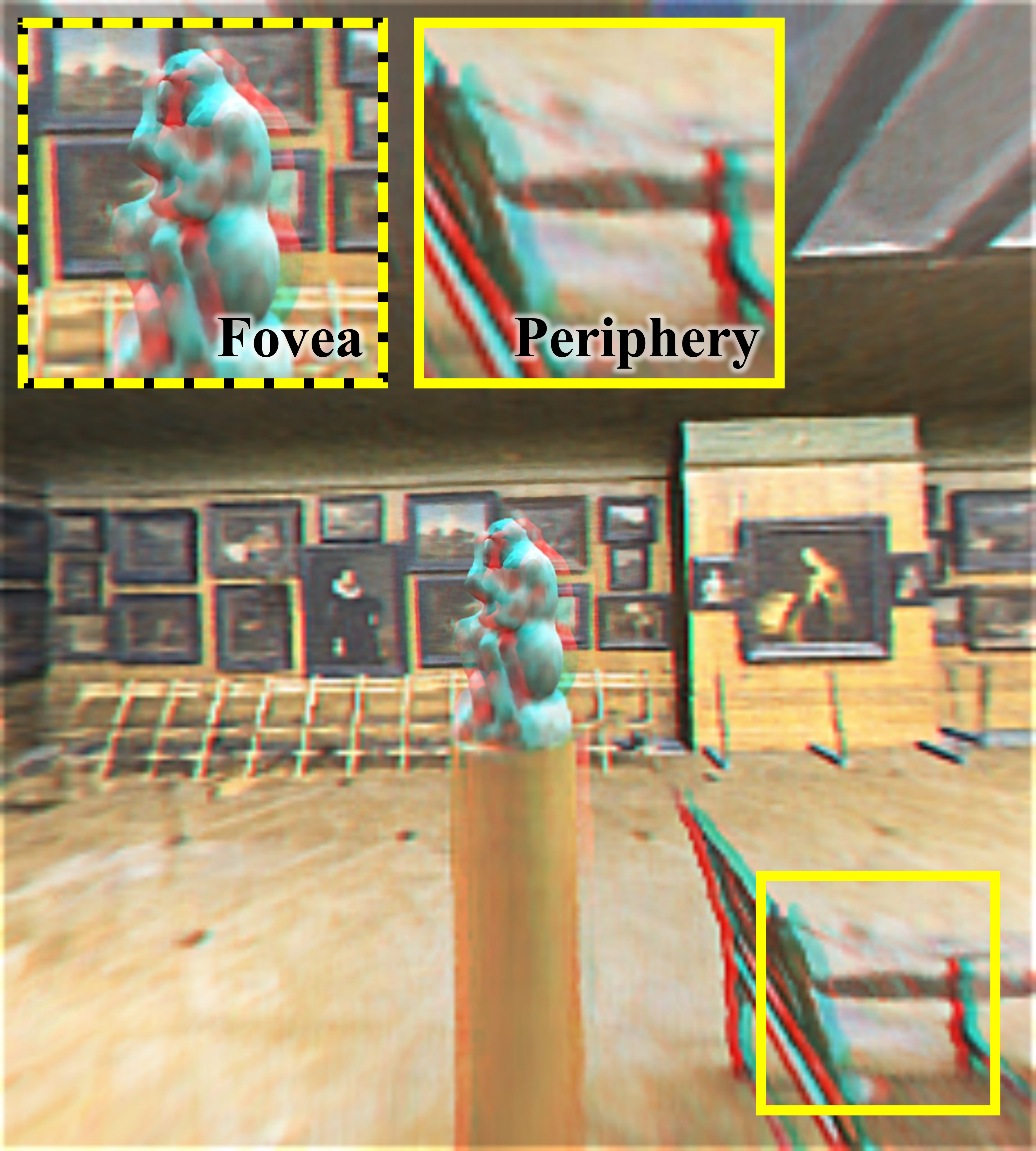}\label{fig:mono:wo}
    }
    \subfloat[\rev{w/ adaptive stereo}]{
        \includegraphics[width=0.48\linewidth]{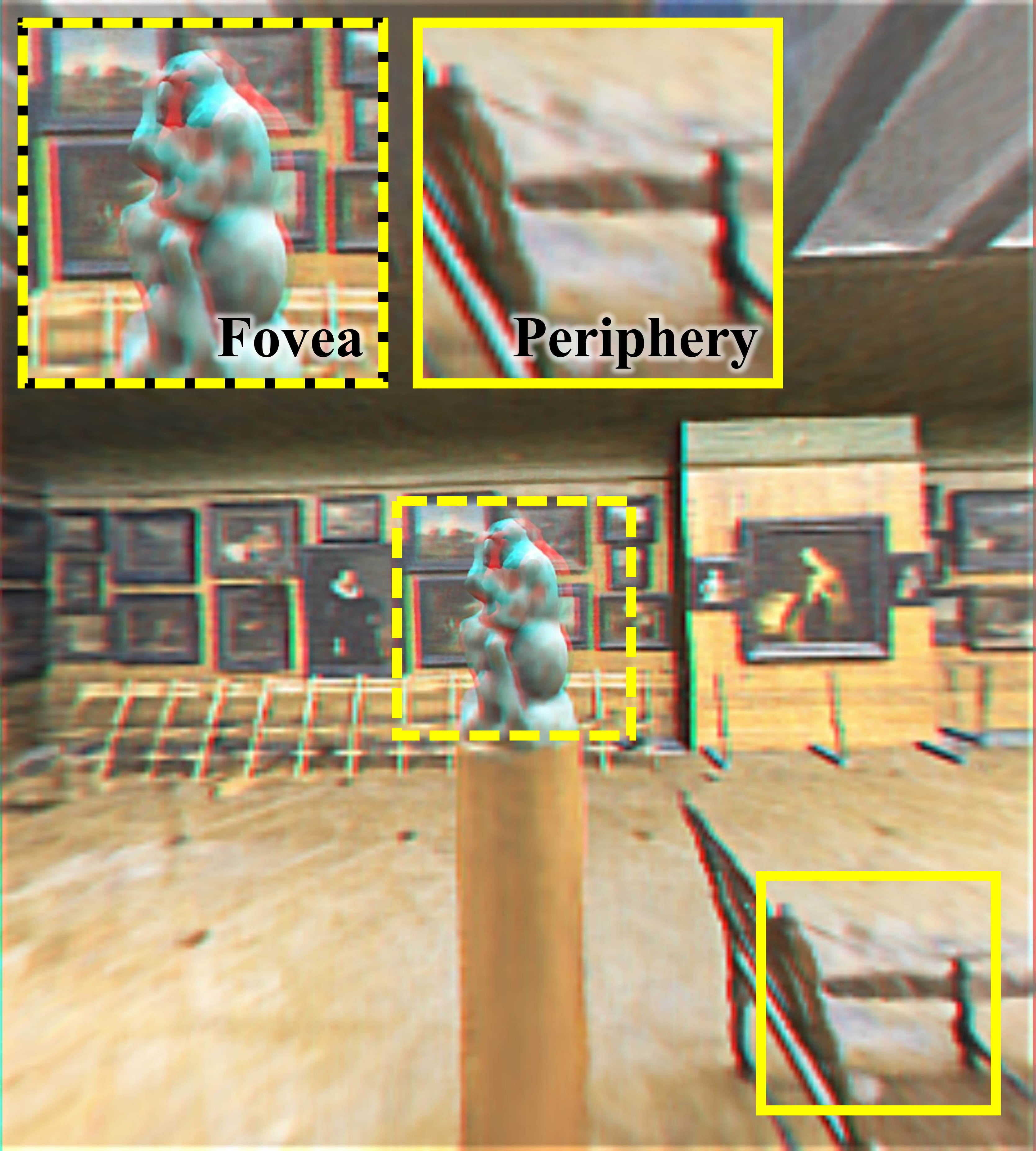}\label{fig:mono:w}
    }

    \Caption{Visualization of the adaptive stereo-acuity with anaglyph.}
    {%
    \subref{fig:mono:wo} shows the rendered image with 6 retinal sub-images ($\imageFoveal^{\{l,r\}}$, $\imageMid^{\{l,r\}}$, and $\imageFar^{\{l,r\}}$).
    \subref{fig:mono:w} shows the rendered image with our adaptive and accelerated inference considering foveated stereoacuity ($\imageFoveal^{\{l,r\}}$, $\imageMid^{\{c\}}$, and $\imageFar^{\{c\}}$). Our method preserves full stereopsis in the fovea while reducing the angular resolution in the periphery for accelerated inference.
    \qisun{I will need to revise this figure.}
    \zh{plan?}
    }
    \label{fig:mono}
\end{figure}

\qisun{(May 15) I feel like this section is a bit thin. Better idea on extending?}
Head-mounted VR displays require stereo rendering to provide parallax depth cues. 
So do the elemental images $\imageFoveal^{\{l,r\}}$, $\imageMid^{\{l,r\}}$, and $\imageFar^{\{l,r\}}$ for each (\underline{$l$}eft and \underline{$r$}ight) eye. 
The stereoscopic rendering, however, doubles the inference computation that is critical for latency- and frame-rate- sensitive VR experience 
\nothing{The separation of the three elemental images considers the varied visual acuity. However, in a head-mounted stereo VR displays, the rendered image are for two eyes, resulting in double computation for $\imageFoveal^{\{l,r\}}$, $\imageMid^{\{l,r\}}$, and $\imageFar^{\{l,r\}}$. Here $l$ and $r$ represent the projection images for left ($\rayo^l$) and right ($\rayo^r$) eyes respectively. 
Whereas, because $\imageMid^{\{l,r\}}$ and $\imageFar^{\{l,r\}}$ demand high spatial resolution due to their large eccentricity coverage, }
(please refer to \Cref{sec:study:intra} for breakdown comparisons).

We accommodate the inference process with the adaptive stereo acuity in perception.
In fact, besides the spatial visual acuity in monoscopic vision, psychophysical studies have also revealed human's significantly declined stereopsis while receding from the gaze point \cite{mochizuki2012magnitude}. Motivated by this characteristic, we perform the computation with $\imageFoveal^{\{l,r\}}$, $\imageMid^c$, and $\imageFar^c$ instead of inferring $6$ elemental images, where $c$ indicates the view at the midpoint of the left and right eyes. \rev{As $\imageMid^c$, and $\imageFar^c$ have zero disparity, obvious misalignment may exist in the blending area of $\imageFoveal^{\{l/r\}}$ and $\imageMid^c$. To reduce this misalignment, we shift $\imageMid^c$ and $\imageFar^c$ according to the vergence of eyes, i.e. the horizontal difference between the left and right gazes, to introduce a disparity that is close to the foveal. Assuming the difference is $\Delta x_g =x_g^l-x_g^r$ (in pixels), then $\imageMid^c$ and $\imageFar^c$ should be shifted by $\Delta x_g/2$ for the left eye and $-\Delta x_g/2$ for the right eye.}
\Cref{fig:mono} visualizes the stereopsis changes from the adaptation using an anaglyph. 

\subsubsection{Real-time frame composition}
With the obtained elemental images as input, an image-based rendering in the fragment shader is then executed to generate final frames for each eye.
The output frames are displayed on the stereo VR HMDs.
Two adjunct layers are blended using a smooth-step function across $40\%$ of the inner layer \rev{as shown in \Cref{fig:system}}.
This enhances visual consistency on the edges between layers \cite{Guenter:2012:F3G}.
\nothing{To accommodate the mono-view $\imageMid^c$ and $\imageFar^c$, they are shifted towards each eye according to approximated foveal depth range.}
Lastly, we enhance the contrast following the mechanism of \cite{Patney:2016:TFR} to further preserve peripheral elemental images' visual fidelity due to its low PDD. 

\subsection{Latency-Quality Joint Optimization}\label{sec:method:optimization}

As a view synthesis system based on sparse egocentric representation (the $\sphereNum$ spheres per network) and neural network synthesis (the $\mlpLayerNum, \mlpChannelNum$), neural rendering methods inevitably introduce approximation errors. \new{The errors can be reduced by introducing additional networks (thus lowered $\sphereNum$ assuming a fixed number of spheres representing a scene) and increasing individual network's capability (i.e., higher $\mlpLayerNum, \mlpChannelNum$). However,}
these variables also significantly increase the online computational time that is determined by inferring function $\mlpFunc$ and ray marching. 
While VR strictly demanding both quality and performance, to seek the optimal latency-quality for human viewers, we present a spatial-temporal model that analytically depicts the correlations and optimizes the variables.

\paragraph{Precision loss of a 3D scene} As shown in \Cref{fig:notations}, under the egocentric representation, a 3D point $\pt$ is re-projected as the nearest point on a sphere that connects it to the origin point:

\begin{equation}\label{eq:closestPoint3D}
{\pt^\prime}(\sphereNum,\mathbf{\sphereRadius},\SpatialPt) \triangleq \sphereRadius_k \frac{\pt}{\norm{\pt}}, \ k=\argmin_{j\in[1,\sphereNum]}\left(\norm{\norm{\SpatialPt}-\sphereRadius_j}\right).
\end{equation}

Similar to volume-based representation, the multi-spherical system is also defined in the discrete domain. The sparsity thus naturally introduces approximation error that compromises the synthesis quality. To analytically model the precision loss, we investigate the geometric relationship among the camera, the scene, and the representation.
As illustrated in \Cref{fig:teaser:scene,fig:notations},  for a sphere (located at origin point) with radius $\sphereRadius$, its intersection (if exists) with a directional ray $\{\rayo,\rayd\}$ is
\begin{equation}
\intersectionFunc(\sphereRadius,\rayo,\rayd) = \rayo + \left({\left((\rayo\cdot\rayd)^2-\norm{\rayo}^2+\sphereRadius^2\right)}^{\frac{1}{2}}-\rayo\cdot\rayd\right)\rayd,
\label{eq:raySphereIntersection}
\end{equation}
where $\rayo$ and $\rayd$ are the ray's origin point and normalized direction, respectively.

Inversely, given a view point $\rayo$ observing \nothing{a spatial location} $\SpatialPt$, the ray connecting them has the direction $\rayd(\SpatialPt,\rayo)=\frac{\SpatialPt-\rayo}{\norm{\SpatialPt-\rayo}}$. This ray may intersect with more than one sphere. Among them, the closest one to $\SpatialPt$ is:
\begin{equation}\label{eq:closestPointImg}
    \hat{\pt}(\rayo,\sphereNum, \mathbf{\sphereRadius},\SpatialPt) \triangleq 
    \intersectionFunc(\sphereRadius_k,\rayo,\rayd(\SpatialPt,\rayo))\ |\ k=\argmin_{j\in[1,\sphereNum]}\left(\norm{\norm{\SpatialPt}-\sphereRadius_j}\right).
\end{equation}
In the 3D space, the offset distance $\norm{\pt^\prime-\hat{\pt}}$ indicates the precision loss at $\pt$ from the representation. By integrating over all views and scene points, we obtain:
\begin{equation}
\begin{aligned}
\sparseError(\sphereNum, \mathbf{\sphereRadius})  &= \iint \norm{\pt^\prime(\sphereNum,\mathbf{\sphereRadius},\SpatialPt)-\hat{\pt}(\rayo,\sphereNum, \mathbf{\sphereRadius},\SpatialPt)} \mathbf{d}\SpatialPt \mathbf{d}\rayo,\\
&\forall \{\rayo, \SpatialPt\} \text{ pair without occlusion in between}.
\label{eq:sparseError}
\end{aligned}
\end{equation}
By integrating all 3D vertices $\SpatialPt$ and camera positions $\rayo$ in our dataset sampling, $\sparseError$ depicts how the generic representation precision of a scene, given a coordinate system defined by $\sphereNum$ and $\mathbf{\sphereRadius}$.

In comparison, a neural representation aims at predicting projected image given a $\rayo$ and $\camDir$. Thus, we further extend \Cref{eq:sparseError} to image space to analyze the error given a set of camera's projection matrix  $\projectionMatrix(\rayo,\camDir)$ as
%
\begin{align}
\imgSpaceError(\sphereNum, \mathbf{\sphereRadius}, \rayo, \camDir)  = \int \norm{\projectionMatrix(\rayo,\camDir)\cdot\left(\SpatialPt-
\hat{\pt}(\rayo,\sphereNum, \mathbf{\sphereRadius},\SpatialPt)\right)}\mathbf{d}\SpatialPt.
\label{eq:imageError}
\end{align}
From \Cref{fig:notations,eq:imageError}, we observe: given a fixed min/max range of $\mathbf{\sphereRadius}$, $\sphereNum$ is negatively correlated to $\imgSpaceError$;
with a fixed $\sphereNum$, the correlation between distribution of $\sphereRadius_j$ and scene content (i.e., distribution of $\SpatialPt$s) also determines $\imgSpaceError$.

However, for neural scene representation, infinitely increasing network capabilities may significantly raise the challenges in training precision and inference performance.
Likewise, increasing representation densities (i.e., lower $\sphereNum$) and/or network complexities (i.e., higher $\mlpLayerNum$/$\mlpChannelNum$) naturally improves the image output quality (lower \Cref{eq:imageError}). However, this significantly increases the computation during ray marching, causing quality drop stretched along time. In the performance-sensitive VR scenario, the latency breaks the continuous viewing experiment and may cause simulator sickness. Thus, with content-aware optimization, we further optimize the system towards an ideal quality-speed balance.

\paragraph{Spatial-temporal modeling} Inspired by \cite{Li:2020:TSP,albert2017latency}, we perform spatial-temporal joint modeling to determine the optimal coordinate system ($\sphereNum$) for positional precision and network complexity ($\mlpLayerNum$, $\mlpChannelNum$) for color precision that adapt to individual computational resources and scene content. This is achieved via latency-precision modeling in the spatial-temporal domain:
\begin{equation}
\begin{aligned}
\finalError(\sphereNum, &\mlpLayerNum, \mlpChannelNum) = 
\sum_t\int \mlpFunc_{\mlpLayerNum, \mlpChannelNum}(\SpatialPt)\times\\ &\norm{\projectionMatrix(\rayo_t,\camDir_t)\cdot\SpatialPt-\projectionMatrix(\rayo_{t-\latency},\camDir_{t-\latency})\cdot\pt(\sphereRadius_k,\rayo_{t-\latency},\rayd_{t-\latency})}\mathbf{d}\SpatialPt,
\label{eq:error:image}
\end{aligned}
\end{equation}
where $\latency\triangleq\latency(\sphereNum, \mlpLayerNum, \mlpChannelNum)$ is the system latency with a given coordinate and network setting.
$\mlpFunc_{\mlpLayerNum, \mlpChannelNum}(\SpatialPt)$ is the four ($r,g,b,a$) output channels' L1-distance between a given network setting and the highest values $\sphereNum=8,\mlpLayerNum=4,\mlpChannelNum=1024$). For simplicity, we assumed uniformly distributed $\mathbf{\sphereRadius}$ with a fixed range of the spherical coverage.

\begin{figure}[tb]
    \centering
    \includegraphics[width=\linewidth]{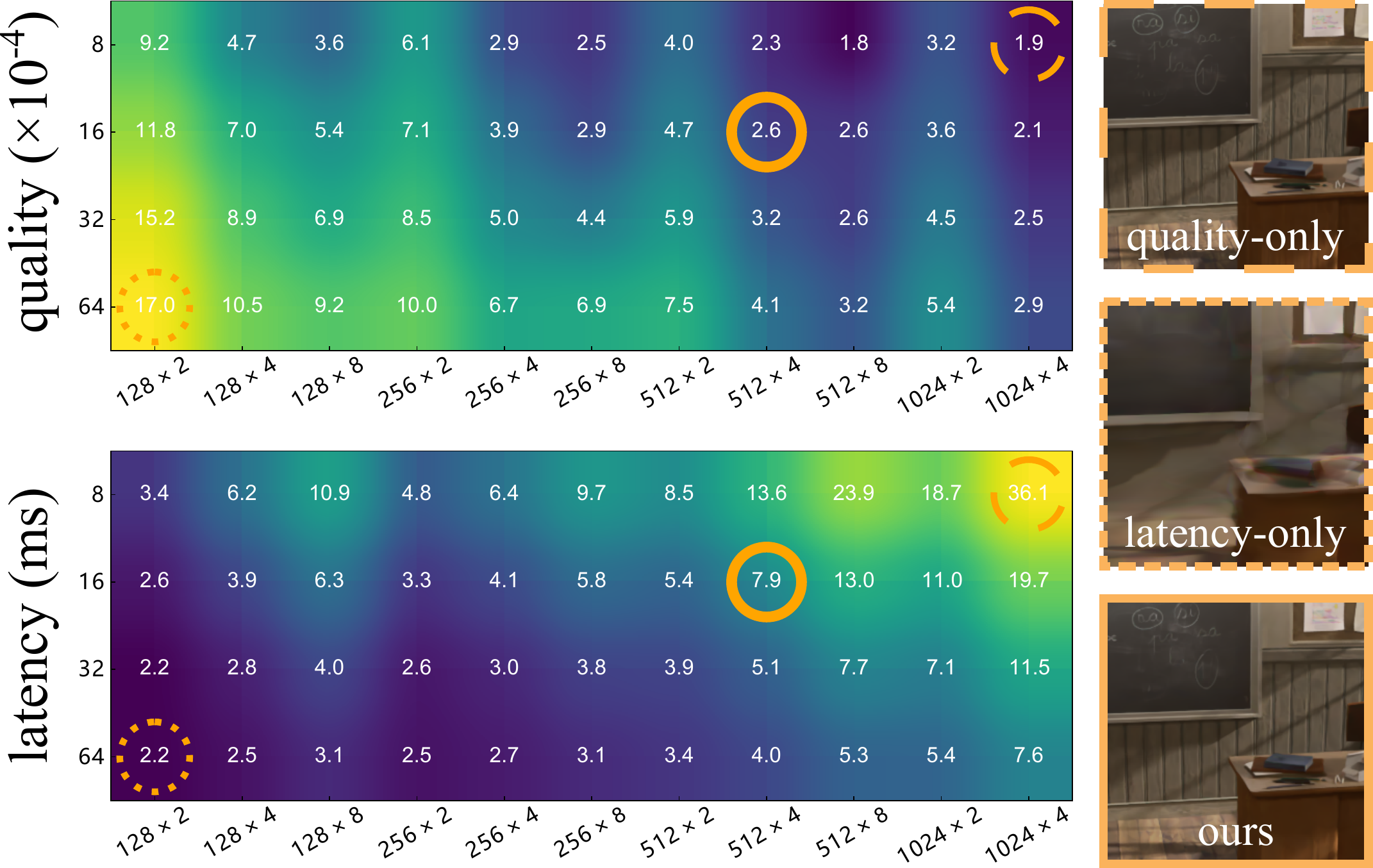}
    
    \Caption{Latency-quality joint-optimization.}
    {%
    The two tables on the left plot $\imgSpaceError$ in \protect\Cref{eq:imageError} and latency of an example foveal network during the optimization process (in milliseconds). The values are computed with various settings of $\mlpChannelNum\times\mlpLayerNum$ (X-axis) and $\sphereNum$ (\new{Y-axis: given a fixed number of spheres representing a scene, lower $\sphereNum$ indicates the need of more networks thus heavier computation}). The images on the right indicate corresponding foveal images under different settings. Our method balances both perceptual quality and latency.
    }
    \label{fig:optimization}
\end{figure}

As suggested by Albert et al. \cite{albert2017latency}, the latency for a foveated system shall reach below \textasciitilde$50$ms for undetectable artifacts. Given our test device's eye-tracking latency \textasciitilde$12ms$ and the photon submission latency \textasciitilde$14ms$ (\cite{albert2017latency}), the synthesis and rendering latency shall be less than $L_0 = 24$ms.
Thus, we determine the optimal $\{\sphereNum, \mathbf{\sphereRadius}\}$ to balance latency and precision as
\begin{equation}
    \argmin_{\sphereNum, \mlpLayerNum, \mlpChannelNum} \finalError(\sphereNum, \mlpLayerNum, \mlpChannelNum), \ \ 
    \text{s.t.}\ {l(\sphereNum, \mathbf{\sphereRadius})} < L_0.
\end{equation}
\Cref{fig:optimization} visualizes an example of the optimization mechanism for an foveal image $\imageFoveal$. The optimized results ($\sphereNum, \mlpLayerNum, \mlpChannelNum$) for individual networks are used for training. The optimization outcomes are detailed in \Cref{sec:impl}.
The visual quality is validated by psychophysical study (\Cref{sec:study:user}) and objective analysis (\Cref{sec:study:quality}). The latency breakdown of our system is reported in \Cref{sec:study:intra}.

\section{Implementation}\label{sec:impl}\zh{TODO}
In this section, we elaborate on how we collect data for training, the specific parameters we choose for the two orthogonal neural networks, and the software/hardware environment we use.

\rev{
\paragraph{Datasets}
We leveraged two groups of datasets: The synthesized datasets (i.e. {\it barbershop}, {\it classroom}, {\it lobby} and {\it stones} used in \Cref{sec:result}) are generated from CG scenes by rendering engine\footnote{{\it barbershop} and {\it classroom} are rendered by Blender's Cycles rendering engine, while {\it lobby} and {\it stones} are rendered through Unity.}. For these scenes, views are sampled uniformly in the translation and rotation box. We generate two separate datasets for every scene to train foveal and periphery networks. The foveal dataset is composed of images rendered with 40 deg field-of-view and $400\times400$ resolution, while the periphery dataset contains images of $400\times400$ with 60 deg field-of-view. Our method is also applicable for physical datasets captured by camera (as illustrated in \Cref{fig:teaser}). For this kind of datasets, we first record a video of a physical scene using mobile phone. Then images are extracted at every second from the video and their poses are evaluated by Colmap\cite{schoenberger2016sfm}. This forms the foveal dataset. The periphery dataset is built by downsample the images in the foveal dataset by factor of 2.
}
\paragraph{Optimized network parameters}
Guided by the latency-quality joint optimization described in \Cref{sec:method:optimization} \new{and $64$ spheres depicting all the tested scenes}, we implemented the foveal network with \new{$\mlpLayerNum=4$, $\mlpChannelNum=512$ and $\sphereNum = 16$}. For the periphery network, the values are \new{$\mlpLayerNum=4$, $\mlpChannelNum=256$ and $\sphereNum = 32$}. These optimized parameters achieve a proper balance between quality and latency, as shown in \Cref{fig:lpips} and \Cref{tbl:ablation}.



\paragraph{Environment}
The system was implemented through OpenGL framework using CUDA and accelerated by TensorRT with Intel(R) Xeon(R) Gold 6230R CPU @ 2.10GHz (256GB RAM) and one NVIDIA GTX 3090 graphics card.
\new{For each scene, we trained both the foveal network and the periphery network with 200 epochs.}
\section{Evaluation}
\label{sec:result}
With various scenes as shown in \Cref{fig:lpips}, we conduct a subjective study (\Cref{sec:study:user}), and calculate objective measurements (\Cref{sec:study:quality}) to evaluate our method's perceptual quality compared with alternative solutions (\cite{mildenhall2020nerf}).
\new{Further, \Cref{sec:study:panorama} validates the method's benefits on enabling large scale translation and view-dependent effects by comparing against prior panorama-based 6DoF synthesize literature \cite{Lin:DeepPanorama}.
}
Lastly, we analyze the intra- and inter-system performance in \Cref{sec:study:intra}.

\subsection{User Study}
\label{sec:study:user}
We conducted a psychophysical experiment to investigate how users perceive our solution ({\bf OURS}) compared with alternative neural view synthesis (\cite{mildenhall2020nerf}, {\bf NeRF}) and full-quality ground truth images ({\bf GT}).


\paragraph{Stimuli}
For precise comparison and accommodating the low frame rate of alternative solutions (\Cref{sec:study:intra}), each group of stimuli consisted of static stereo images rendered via {\bf GT} or synthesized via {\bf OURS}/{\bf NeRF} with the same views, as shown in \Cref{fig:lpips}.
They were generated with the same and randomly defined gaze fixation across conditions.
The resolution of the image per eye was $1440 \times 1600$ (full capability of the VR display). 
During the study, the target gaze position was indicated as a green cross on the stimuli images.
We used the {\it classroom}, {\it lobby} and {\it barbershop} scenes for study use.
Each condition from an individual scene consisted of \nothing{$1$ view and} $2$ different gaze positions.

For generating {\bf NeRF} condition, we retrained the model from \cite{mildenhall2020nerf} on our dataset with $\mlpLayerNum=8$, $\mlpChannelNum=256$, and $\sphereNum = 64$ for coarse network and $\sphereNum = 128$ for fine network. The top right insets of \Cref{fig:lpips} show the resulting {\bf NeRF} images. 


\paragraph{Setup}
Each participant wore an eye-tracked HTC Vive Pro Eye headset and remained seated to examine the stimuli during the experiment.
Twelve users participated in and completed the study ($6$ females, $M=23.25$\nothing{, $SD=2.14$}). 
No participants were aware of the research, the experimental hypothesis, nor the number of conditions. All participants had a normal or corrected-to-normal vision.

\paragraph{Task}
The task was a \textit{two-alternative-forced-choice} (2AFC). 
Each trial consists of a pair of stimuli generated from two of the three methods ({\bf OURS} / {\bf NeRF} / {\bf GT}) with a sampled view and gaze position. 
\new{
To avoid gaze motion variances among individual trials, we enforced static gaze than free-form viewing in each pair of trials.
Each stimulus appeared for $500$ms on the display. The duration was designed to prevent refocusing and unintentional shifting (0.3-0.5s) as studied in \cite{campbell1960dynamics}.
}
A forced $0.3$sec break (black screen) was introduced between conditions to flush the vision. 
During the study, the participants were instructed to fix their gazes on a green cross rendered on the display. 
To prevent the fixation from shifting away and ensure accuracy, we tracked the users' gaze throughout the experiment. Whenever the gaze is more than $5$ deg away from the target, a trial was dropped immediately with a black screen informing the participant.
After each trial, the participants were instructed to select which of the two stimuli appeared with higher visual quality using a keyboard.
Before each experiment, a warm-up session with 6 trials was provided to familiarize the participants with the study procedure. The orders of conditions among trials were randomized and counter-balanced. The entire experiment of each user consists of $36$ trials, $6$ trials per pair of ordered conditions.
To minimize the effect of accumulated fatigue, we also enforced breaks between trials (at least 2 seconds) and after each scene (60 seconds). Meanwhile, the participants were allowed to take as much time as needed. 

\begin{figure}[tb]
    \centering
    \includegraphics[width=0.96\linewidth]{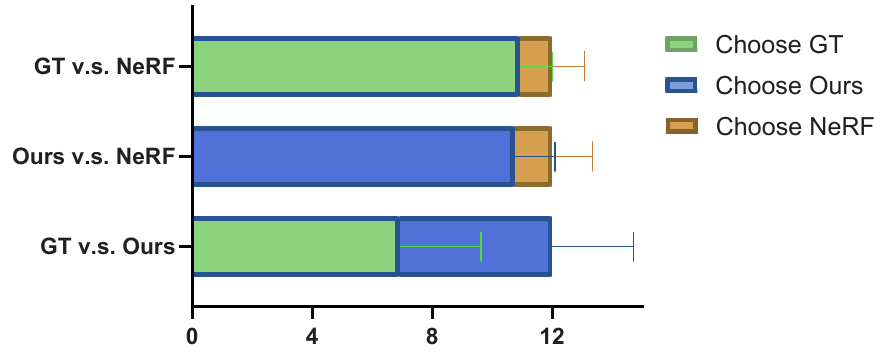}
    \Caption{The users' preference votes from our evaluation experiment (\protect\Cref{sec:study:user}).}
    {%
    X-axis shows stacked vote results of selecting the condition of each pair. Y-axis lists the pair accordingly.
    }
    \label{fig:results:2afc}
\end{figure}

\paragraph{Results}
\Cref{fig:results:2afc} visualizes the results considering all conditions and their orders in the 2AFC experiments. Here we analyzed and reported the results regardless of the orders.
Among all three conditions, we observed close-to-random-guess among trials that compared {\bf GT} and {\bf OURS} ($42.4\%$ voted for {\bf OURS}, $SD=0.23$). Meanwhile, a significantly higher ratio of voting {\bf GT} over {\bf NeRF} was observed ($91.0\%$, voted for {\bf GT}, binomial test showed $p<0.005^{***}$, $SD=0.11$). The preference applies to {\bf OURS} vs. {\bf NeRF} as well ($89.6\%$ voted for {\bf OURS}, $SD=0.09$, binomial test showed $p<0.005^{***}$). \new{We then calculated the effect size when sample size is 12 (in our case) and set power to 0.8. To reach 0.8 for power value, effect size is at least to be 1.69 and our effect size $Cohen's d = 3.5$ for Ours v.s. NeRF, and $Cohen's d = 4.5$ for GT v.s. NeRF. }

\paragraph{Discussion}
The close-to-random-guess ($50\%$ given a 2AFC task) results in {\bf OURS} vs. {\bf GT} revealed the statistical perceptual similarity. 
Meanwhile, both conditions showed significant quality preference than {\bf NeRF}. That is, with immersive, stereo, high-FoV, and egocentric viewing settings, {\bf OURS} synthesizes gaze-contingent retinal images with superior perceptual quality than alternative  solutions (\cite{mildenhall2020nerf}).
Given a fixed network representation capability, {\bf NeRF} synthesizes the whole visual field with identical quality while {\bf OURS} concentrates the representation based on the optimal spatial and stereoscopic acuity.



\subsection{Static Visual Quality}
\label{sec:study:quality}
\paragraph{Conditions and metrics}
Complementary to the subjective measurement (\Cref{sec:study:user}), we further objectively validate the perceptual quality at individual eccentricities in a breakdown fashion.
Specifically, with all four scenes and {\bf GT} as the quality reference, we compare 
{\bf OURS}, 
{\bf NeRF},
and an additional image-space spatial foveation {\bf F-GT} \cite{jiang2015salicon}.
As has been studied to be perceptually identical to {\bf GT}, the {\bf F-GT} condition serves as a spatially adaptive quality reference.

For each eccentricity range, we compare the deep perceptual similarity (LPIPS) \cite{zhang2018unreasonable} across all scenes. LPIPS uses deep neural networks to estimate perceptual similarities between the image provided and a reference image. \new{It widely serves as image-based metrics\nothing{ that a perceptual pairwise image distance is calculated to} that quantifies humans' visual similarity judgments. Smaller values indicate higher perceptual similarity. Mildenhall et al. \cite{mildenhall2020nerf} also applied LPIPS for quality measurement.}
For each scene, we sample $25$ views with gazes at the middle of the display, resulting in $25$ data per eccentricity value, $21$ eccentricity value per scene ($5$ deg step size), thus $525$ data per scene. We used one-way repeated measures ANOVAs to compare effects across three stimuli on LPIPS value of each eccentricity range (IV: stimuli, DV: LPIPS value). Paired t-tests with Holm correction were used for all pairwise comparisons between stimuli. All tests for significance were made at the $\alpha=0.05$ level. 
The measurement was conducted at individual eccentricity ranges (from $0$ to $110$ deg, the capability of the VR HMD). 

\begin{figure*}[htbp]
    \centering
    \newcommand{\figH}{0.128}
    
    \subfloat[classroom analysis]{\label{fig:lpips:classroom}\includegraphics[height=\figH\paperheight]{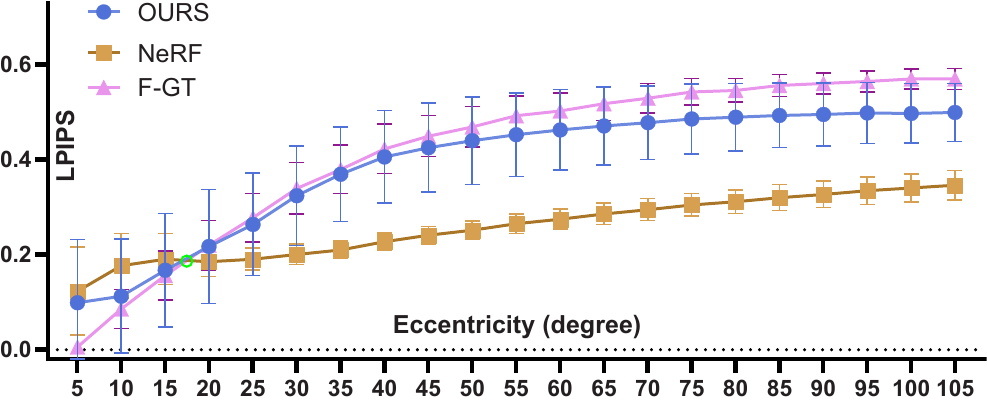}}
    \subfloat[classroom example]{\includegraphics[height=\figH\paperheight]{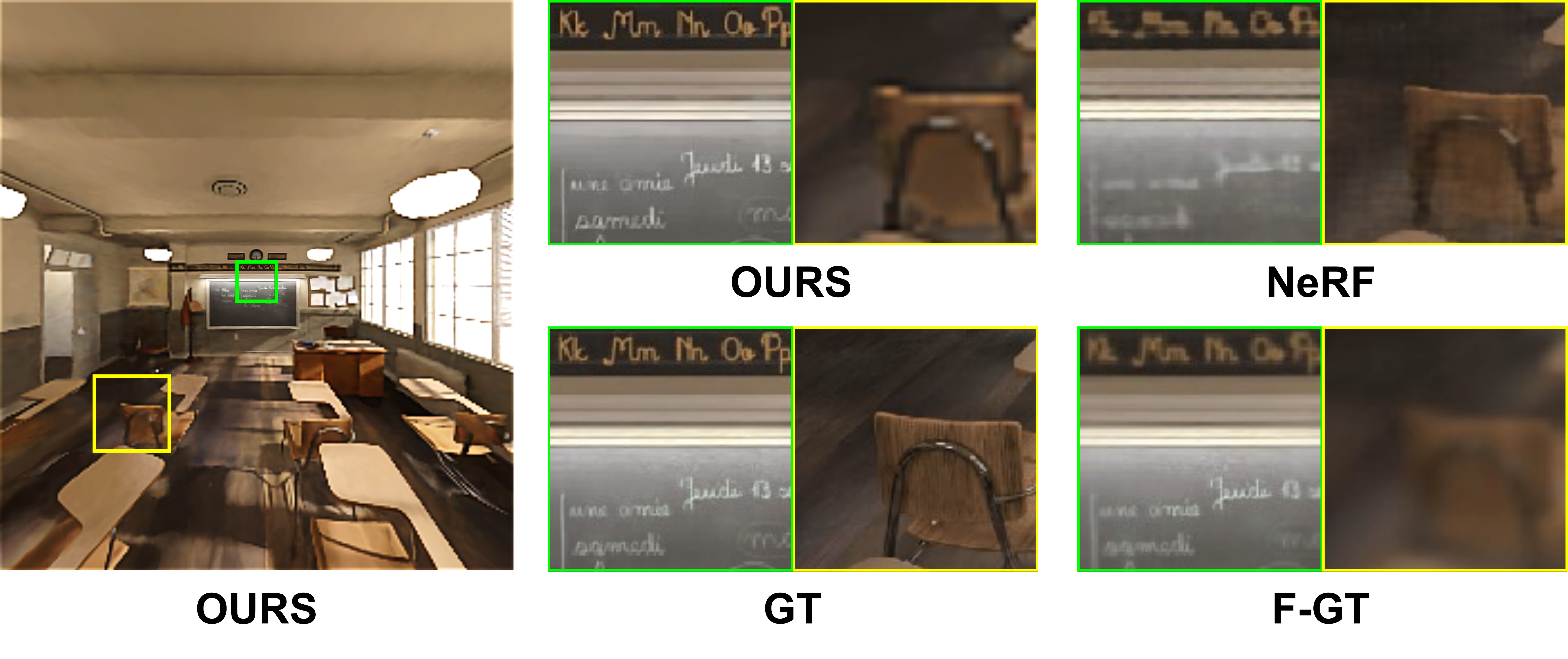}}\label{fig:results:classroom}
    
    \subfloat[lobby analysis]{\label{fig:lpips:lobby}\includegraphics[height=\figH\paperheight]{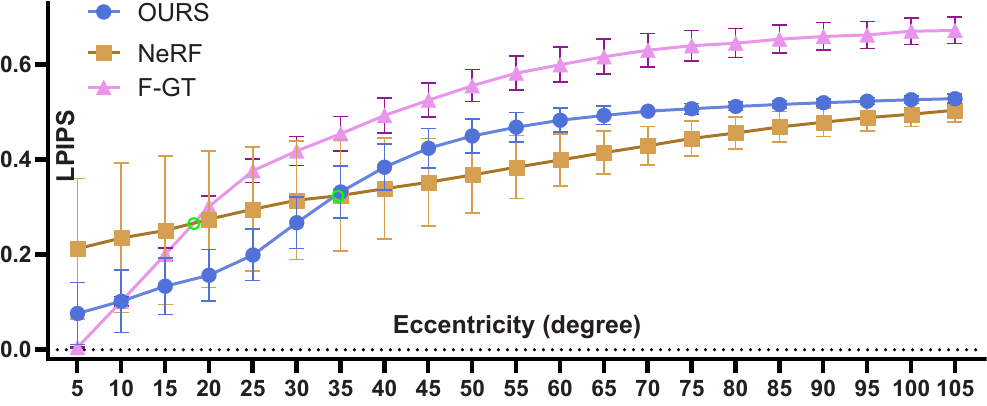}}
     \subfloat[lobby example]{\includegraphics[height=\figH\paperheight]{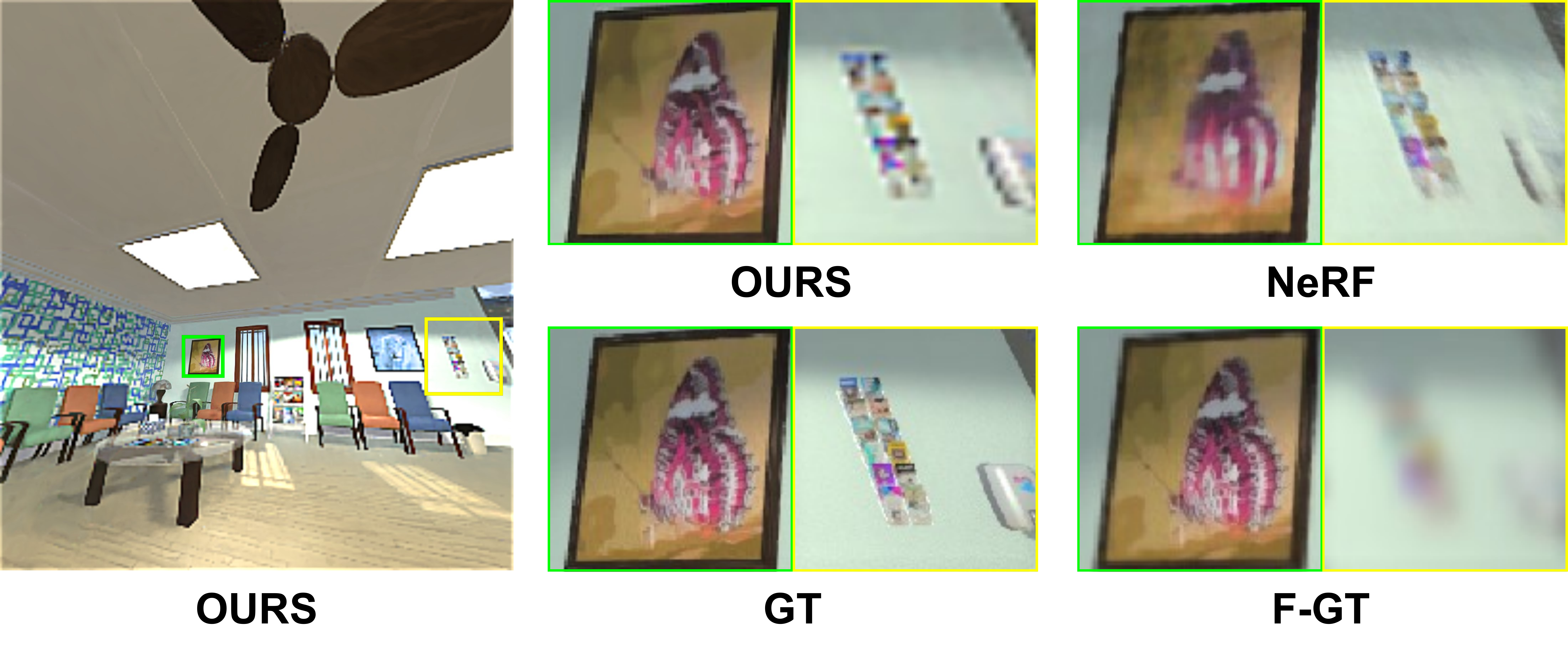}}\label{fig:results:lobby}
    
    \subfloat[stone analysis]{\label{fig:lpips:stone}\includegraphics[height=\figH\paperheight]{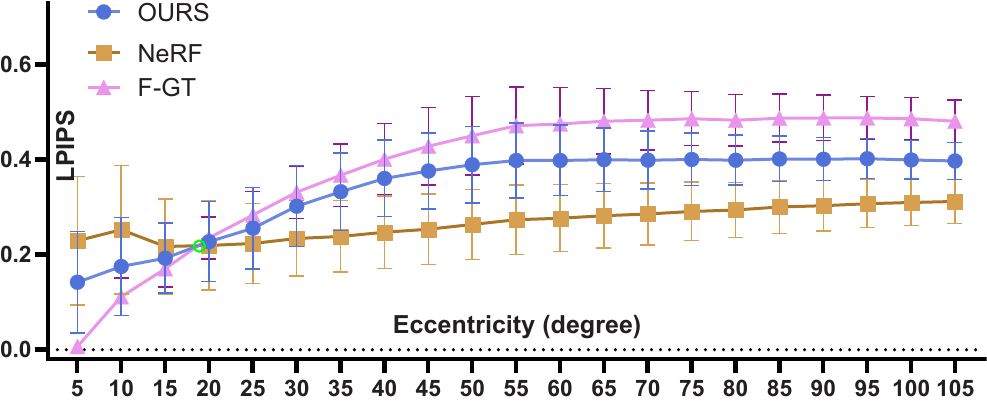}}
    \subfloat[stones example]{\includegraphics[height=\figH\paperheight]{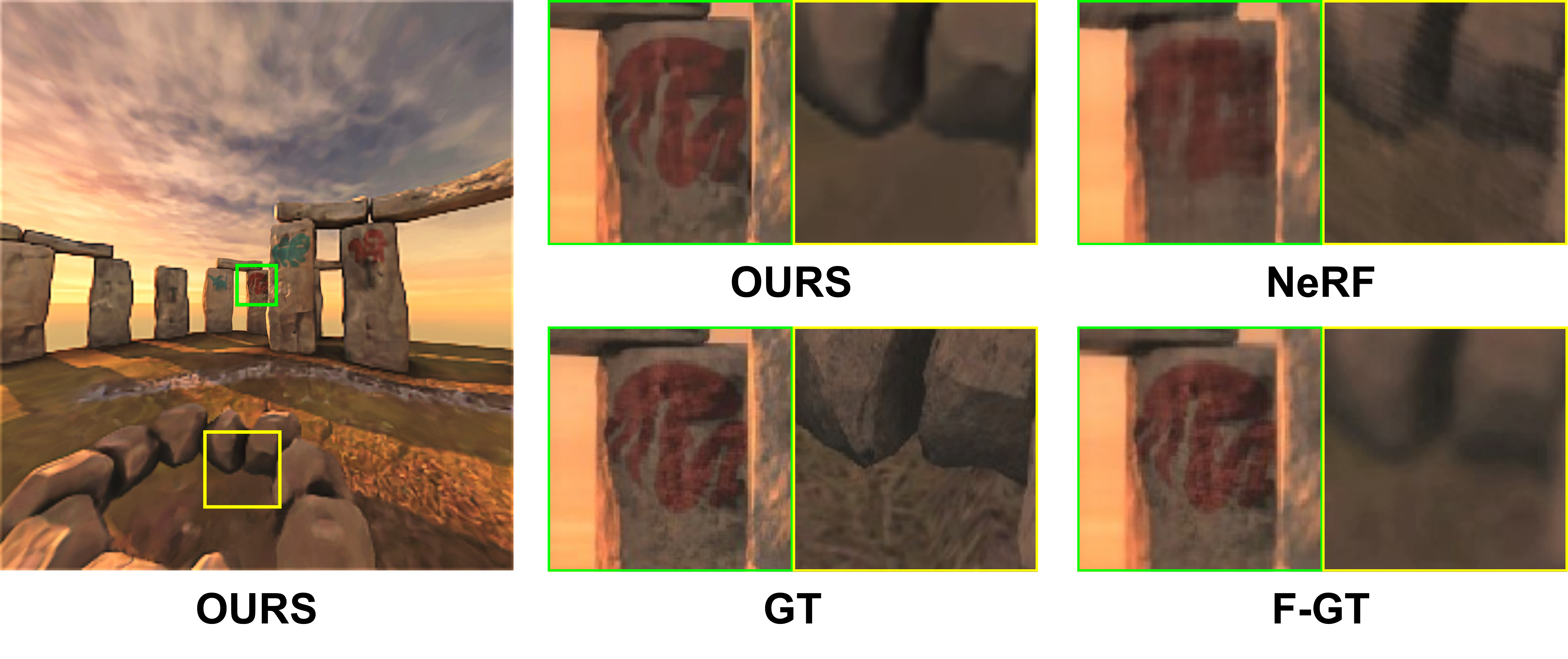}}\label{fig:results:stone}
    
    \subfloat[barbershop analysis]{\label{fig:lpips:barbershop}\includegraphics[height=\figH\paperheight]{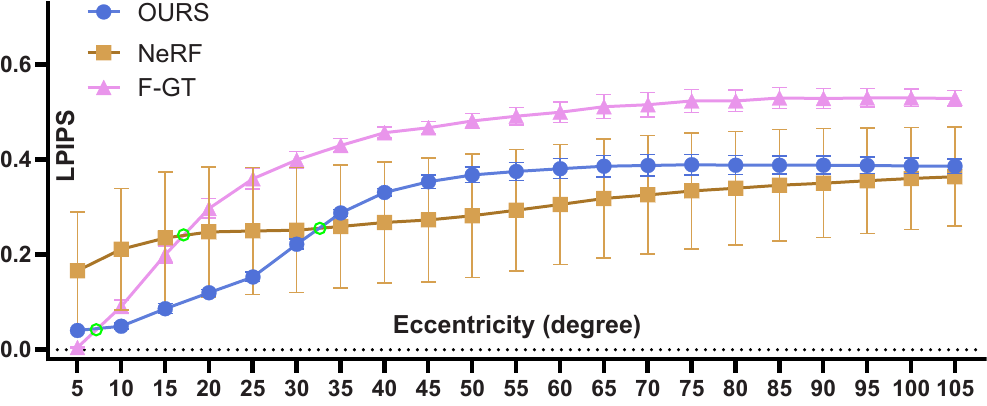}}
    \subfloat[barbershop example]{\includegraphics[height=\figH\paperheight]{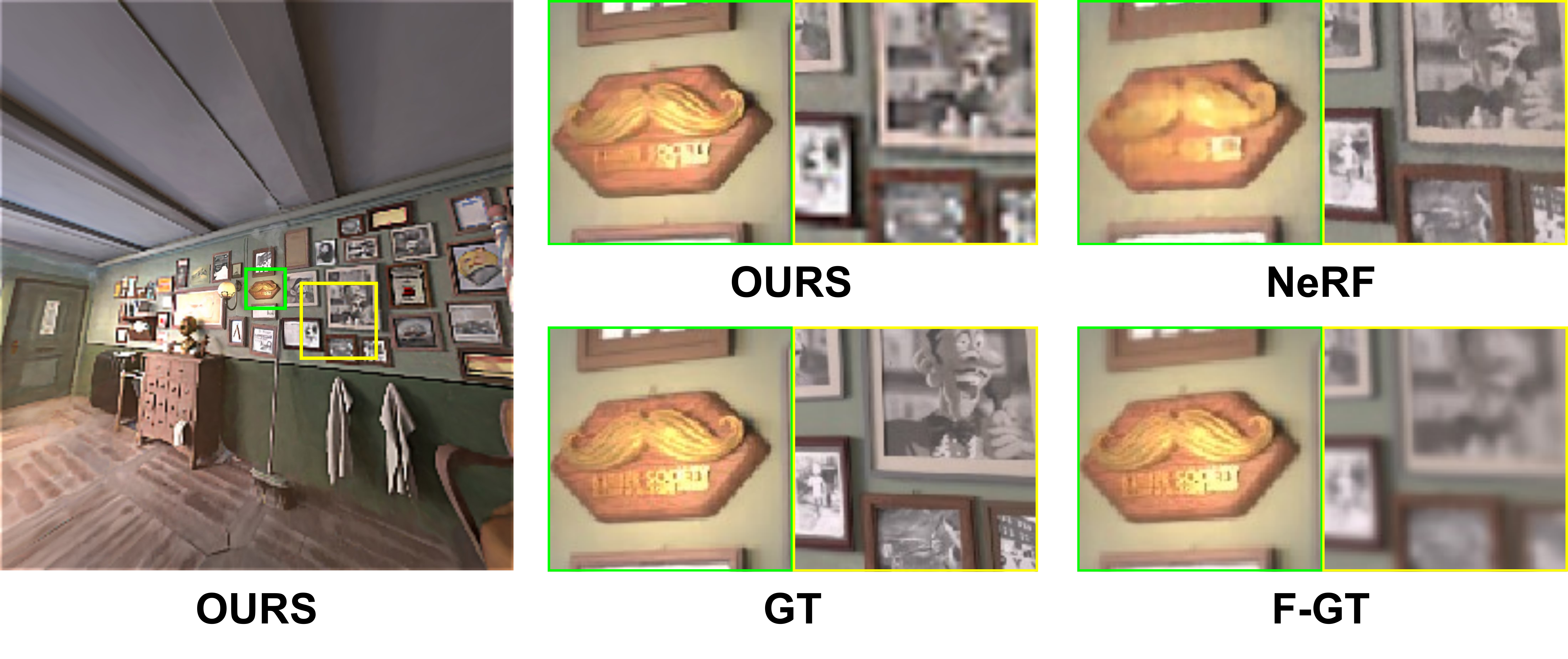}}\label{fig:results:barbershop}
    
    \Caption{LPIPS visualization of all scenes and comparison of {\bf OURS}, {\bf NeRF}, and {\bf F-GT} over the visual field.}
    {%
    The left column plots the analysis results from \protect\Cref{sec:study:quality}. 
    X-axis indicates the eccentricity range ($0$ to $110$deg). Y-axis shows the LPIPS loss \cite{zhang2018unreasonable} with {\bf GT} condition as reference. Lower values mean more similar perceptual quality to the reference, i.e., better quality. The error bars indicate the standard deviation. The green circles indicate intersection points between curves.
    The right column visually compares foveal/periphery images with randomly sampled views. Green/yellow rectangle indicates fovea/periphery.
    \warning{Redo the caption and figure legends when we finish!}
    }
    \label{fig:lpips}
\end{figure*}

\paragraph{Results} 
\Cref{fig:lpips} plots LPIPS values across all scenes and eccentricity ranges ($5$ deg step size). 
From foveal to near periphery ($\leq15$ deg), we observed significant effects of the stimuli on LPIPS with a ``large'' effect size ($\eta^2 >= 0.15$). That is, {\bf OURS} shows significantly lower LPIPS than {\bf NeRF} ($p<=.005^{***}$), and being higher than {\bf F-GT} in scene classroom and lobby or lower than {\bf F-GT} in scene barbershop and stones.
For example, the main effects of stimuli ($F(2,48)=24.74, p<.001^{***}$) was significant on eccentricity $=[0,30]$ deg in scene \textit{barbershop} (\Cref{fig:lpips:classroom}). {\bf OURS} was significantly lower than {\bf NeRF} ($t(24)=-4.26, p<.001^{***}$) and {\bf F-GT} ($t(24)=-9.1, p<.001^{***}$) both with a ``large'' effect size (Cohen's d $>0.85$). The example observation generally applies to all 4 scenes being validated. 
The trend extends to ($\leq35$ deg) for the ``lobby'' and ``barbershop'' scenes.\qisun{@Zhenyi: add a significance analysis to validate this claim here.} \zh{don't understand}

From near- to far- eccentricity (>$20$ deg), we observed significant effects of the stimuli on LPIPS with a ``large'' effect size ($\omega^2 = 0.32$). {\bf OURS} shows higher LPIPS than {\bf NeRF} and it became significantly from eccentricity=$40$ deg. Whereas, comparing with {\bf F-GT}, we observed significant lower scores ($p<.001^{***}$).
For instance, the main effects of stimuli was significant on eccentricity $=40$ deg in scene \textit{lobby} ($F(2,48)=87.78, p<.001^{***}$, \Cref{fig:lpips:lobby}). {\bf OURS} was significantly lower than {\bf F-GT} ($t(24)=-9.095, p<.001^{***}$), and higher than {\bf NeRF} ($t(24)=3.797, p<.001^{***}$) both with a ``large'' effect size (Cohen's d $=0.76$). The example observation generally applies to all 4 scenes being validated.

\paragraph{Discussion}
In the foveal and near-periphery, the observation revealed our method's significantly higher visual quality than the alternative solutions. This is evidenced by the significantly stronger perceptual similarity to {\bf GT} by comparison between {\bf OURS} and {\bf NeRF}. 
In our subjective study (\Cref{sec:study:user}), the statistically indistinguishable/preferred voting of {\bf OURS} vs. {\bf GT}/{\bf NeRF} also agrees with the discovery.
In the periphery with eccentricity larger than $20$ deg, {\bf OURS} showed lower LPIPS (i.e., perceptually more similar to {\bf GT}) than {\bf F-GT}.
The latter has been shown to display identical perceptual similarity to full resolution rendering \cite{Patney:2016:TFR}.
Thus, the findings show that {\bf OURS} doesn't compromise the peripheral vision's quality with its significantly enhanced synthesis acuity in the fovea. 
Note that the discoveries above also agree with the observations from \Cref{sec:study:user}. That is, in addition to the significantly faster performance ($99.8\%$ time reduction per frame as in \Cref{sec:study:intra}), our method showed superior perceptual quality than {\bf NeRF} under first-person, high resolution, and immersive viewing. 
\new{
\subsection{Dynamic Viewing Quality}
We further validated our method on dynamic scenes allowing free head and gaze motions in a VR immersive environment.
Subjective 2AFC evaluations requiring free gaze motion are limited by a fundamental problem where the gaze trajectories differ between discrete 2AFC trials. This inevitably results in variations in perceived stimuli between any two methods being compared, thereby resulting in erroneous comparisons arising from potential bias. To overcome this problem, we instead employ a novel combination of {\it{subjective}} viewing trajectories and recent {\it{objective}} perceptual metric for foveated video quality.

Specifically, we collect subjective gaze data wherein the users were asked to freely observe the full-resolution ground truth ({\bf GT}) immersive VR video. We then generate foveated videos using state-of-the-art foveated rendering methods to compare against our method with the collected gaze and head data as the baseline. This way, we make sure that the gaze trajectories across all the methods are maintained same. To predict the perceived visual quality of video stimuli generated using different methods, we use the most recent FovVideoVDP metric~\cite{mantiuk2021fovvideovdp} which is tailored for large field-of-view (e.g., VR) video sequences considering various perceptual factors such as spatial and temporal effects across visual eccentricities, motion and contrast.
The metric effectively predicts the levels of visible visual differences between a given video stimulus and a reference. 

\paragraph{Conditions and metrics}
In this experiment, we collected 12 randomly sampled (gaze and head) trajectories from the participants in \Cref{sec:study:quality}. Each trajectory forms one trial that is about 2 seconds long (i.e., 100 frames with {\bf OURS}). 
Due to FovVideoVDP's modeling of only monocular views, we measure with the rendered frames with left eyes only, without the loss of generality.
{\bf GT}, the video sequence rendered with full-resolution, was considered as the reference in FovVideoVDP.
With the target reference, we compare our method with various gaze-contingent methods, including the recently presented real-time ventral metamers \cite{walton2021beyond,freeman2011metamers} ({\bf M}(etamer){\bf-GT}) and foveated real-time foveated rendering \cite{jiang2015salicon} ({\bf F-GT}). Due to the significantly low temporal performance of {\bf NeRF} condition (see \Cref{sec:study:intra}), the trial duration (2 sec) is shorter than the needed time to generate a single frame with it (90 sec). It showed a significantly low FovVideoVDP rate ($\leq 5.1$). Thus, we exclude the unnecessary statistical comparison with this condition.

\paragraph{Results and discussion}
\Cref{fig:results:fvvdp1} and \Cref{tab:results:fvvdp} show the results' statistical distribution. Specifically, with FovVideoVDP's visual similarity range from $0$ to $10$, {\bf OURS} achieves $M=8.26, SD=0.14$, significantly higher than {\bf M-GT} $M=6.77, SD=0.18$ and {\bf F-GT} $M=7.99, SD=0.19$ with $F(11, 22) = 14.15, p<.001^{***}$. 
We further perform a power analysis by setting sample size to 12 and power to 0.8. To reach 0.8 for power value, effect size is at least to be 1.69 and our effect size for {\bf OURS} v.s. {\bf F-GT} is $Cohen's d = 2.58$, and $Cohen's d = 14.24$ for {\bf OURS} v.s. {\bf M-GT}. 

\begin{figure}[tb]
    \centering
    \includegraphics[width=0.9\linewidth]{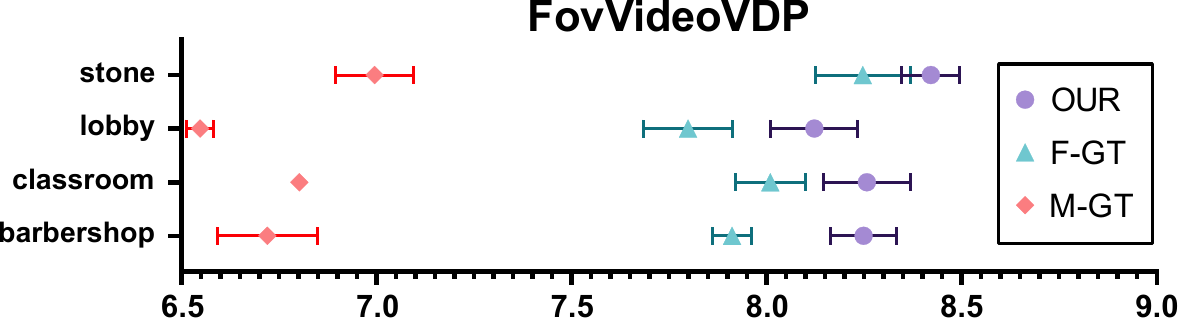}
    
    \Caption{The descriptive results from fvvdp measurement.}
    {%
    X-axis shows the FovVideoVDP results (higher values mean perceptually closer to the reference {\bf GT} condition). Y-axis lists four scenes being validated. Error bars indicate standard deviation.
    }
    \label{fig:results:fvvdp1}
\end{figure}

\begin{table}[tb]
    \centering
    \begin{tabular}{cccc}
         Comparison & Mean Diff & 95\% CI of Diff & P Value   \\
         \midrule
         OURS vs F-GT & 0.2706 & 0.1861 to 0.3551 & $<.0001^{****}$ \\
         OURS vs M-GT & 1.496 & 1.356 to 1.635 & $<.0001^{****}$ \\
         F-GT vs M-GT & 1.225 & 1.144 to 1.307 & $<.0001^{****}$ \\
    \end{tabular}
    \caption{The post-hoc comparison results for FovVideoVDP ANOVA.}
    \label{tab:results:fvvdp}
\end{table}

In addition to the previous experiment on controlled static viewing (\Cref{sec:study:quality}), the analysis demonstrates our outperformance in spatio-temporal visual quality under dynamic viewing conditions. 
This is validated via comparing with existing rendering approaches that have been demonstrated to provide perceptually identical quality to a full resolution rendering (i.e., {\bf GT} the reference of the metric). The experiment simulates real-world viewing scenarios in immersive environments where users view constantly moving with their heads and gazes.
}
\subsection{Qualitative Comparison with Panorama-Based View Synthesis}
\label{sec:study:panorama}
\paragraph{Conditions and implementation}
The synthesis quality of panoramic-imagery-based approaches is determined by the occlusion and translation ranges. 
Thus, as a case study,
we visually compare our synthesis approach with \cite{Lin:DeepPanorama} and varied
translation ranges from the camera origin. 
To produce a \emph{Layered Depth Panorama} (\cite{Lin:DeepPanorama}, {\bf PANO}), we render $16$ images from a virtual camera ring rig around the origin of the scene. Then as an identical input format, we passed the images to
Lin et al.'s  pre-trained model. 
Once a {\bf PANO} is generated, we query it for images corresponding to the
same camera poses used for {\bf OURS}. 

\begin{figure}[tb]
    \subfloat[large translation w/ occlusion]{\includegraphics[width=0.47\linewidth]{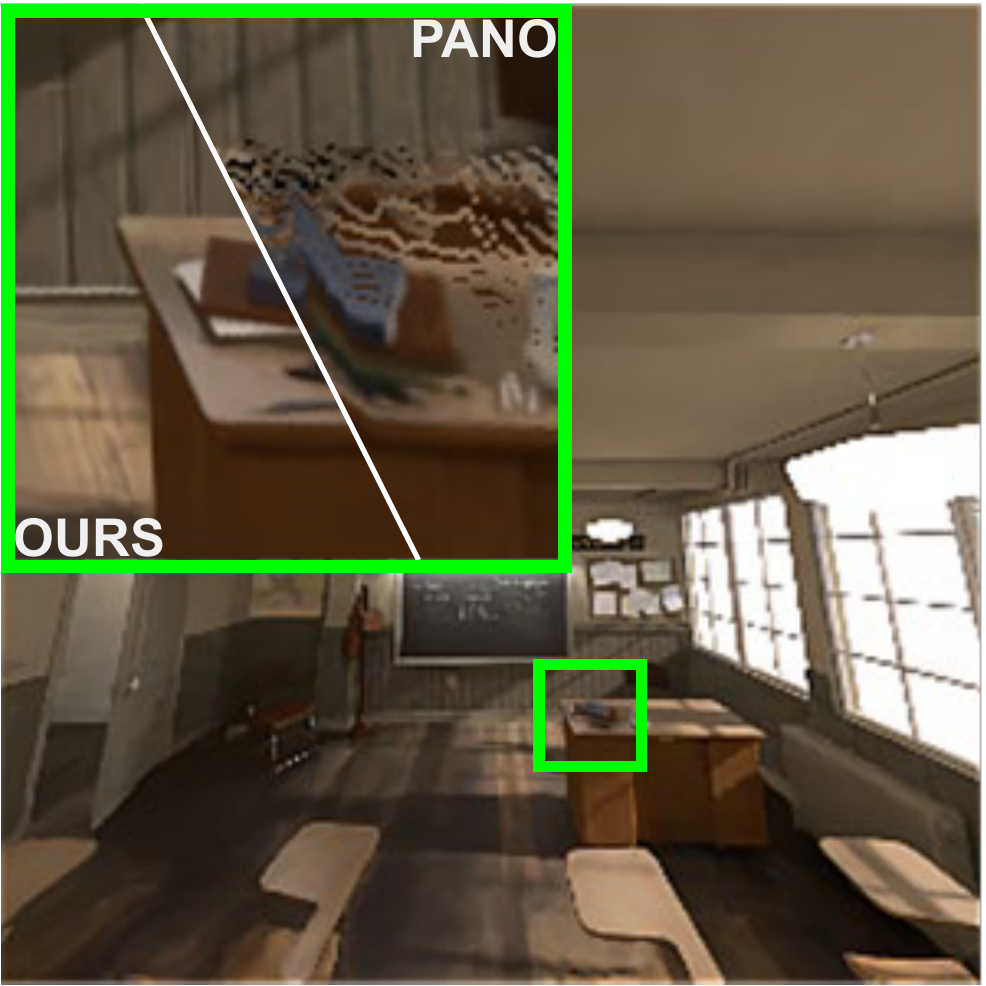}\label{fig:panorama:large}}       
    \subfloat[small translation w/ flat surface]{\includegraphics[width=0.47\linewidth]{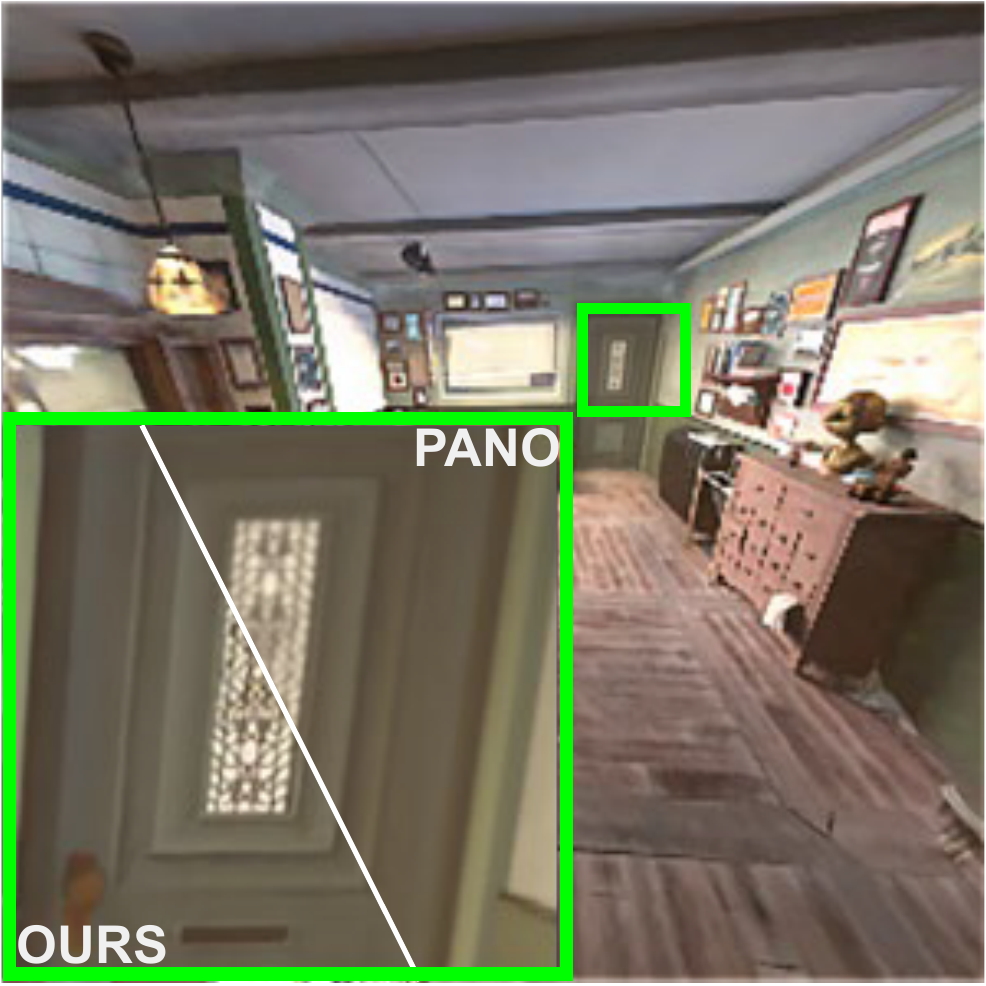}\label{fig:panorama:small}}
    
    \Caption{Qualitative comparison between {\bf OURS} and panorama-based view synthesis.}
    {
    Comparison between our model ({\bf OURS}), and {\bf PANO}\cite{Lin:DeepPanorama}
    shows that the quality of images varies most where the scene contains complex
    geometries, and/or a wide scale of depth variance and occlusion (cf.~\emph{desk with
    books} in \subref{fig:panorama:large}). Meanwhile, presence of high frequency textures (cf.~\emph{door window} in \subref{fig:panorama:small}) show
    marginal difference between the methods.
    }
    \label{fig:panorama}
\end{figure}

\paragraph{Dis-occlusion quality}
\Cref{fig:panorama} shows sampled views of failure (large translation range of 30cm while facing occluded geometries, \Cref{fig:panorama:large}) and success (small translation range of 15cm while facing a flat surface, \Cref{fig:panorama:small}) modes of {\bf PANO} in comparison to {\bf OURS}. 
We observe that our model performs better in scenes with high occlusion and depth variance, while presenting a marginal difference for mostly flat geometries. 
We hypothesize that this observation is largely because our approach presents better congruence towards learning complex geometries, whereas {\bf PANO} is biased towards modeling geometries as planes. We note that this bias is likely conducive to {\bf PANO}'s ability to retain high frequency textures if the variance in depth is minimal. \monde{not sure how
to justify this claim.}

\begin{figure}[tb]
    \centering
    \subfloat[\textbf{OURS}]{\includegraphics[width=0.96\linewidth]{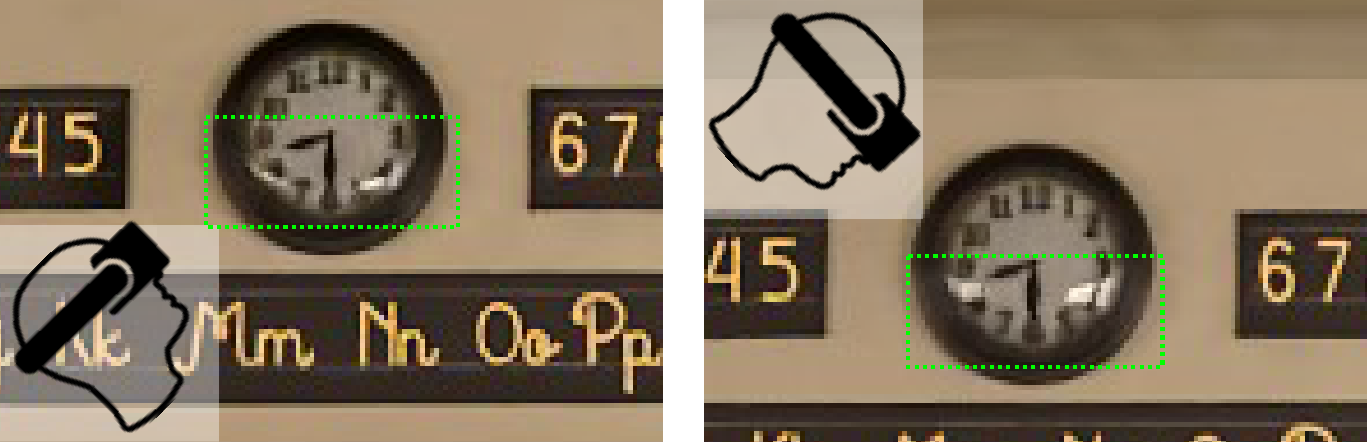}\label{fig:vd:our}}
    
    \subfloat[\textbf{PANO}]{\includegraphics[width=0.96\linewidth]{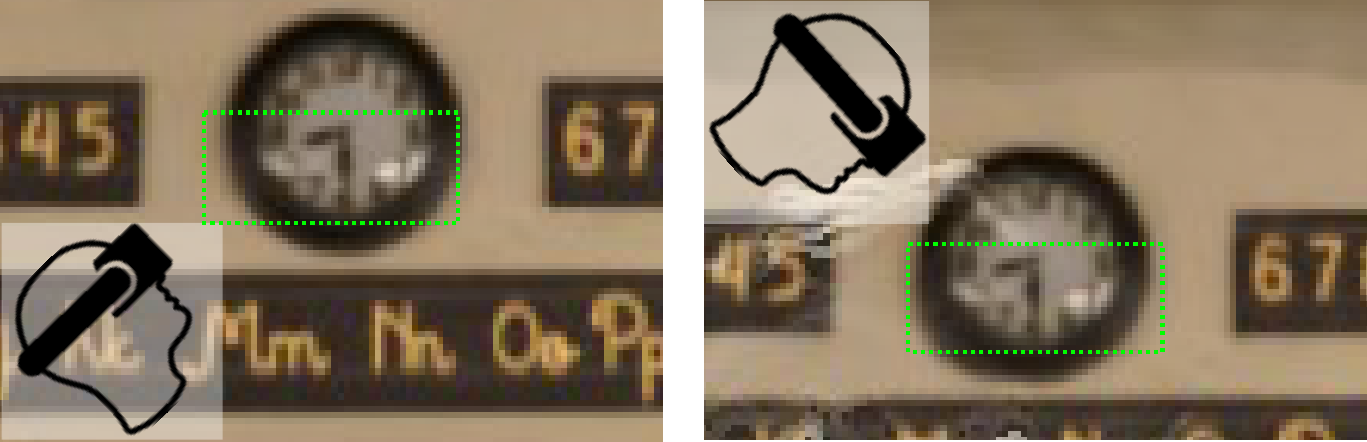}\label{fig:vd:pano}}
    \Caption{Comparing view-dependent effects between \textbf{OURS} and \textbf{PANO}.} 
    {%
    Note the view-dependent effect appeared on the glossy wall-mounted clock.
    With an up-down rotating and translating camera, \textbf{OURS} \subref{fig:vd:our} shows stronger view-dependent effects than \textbf{PANO} \subref{fig:vd:pano} condition.
    }
    \label{fig:vd}
\end{figure}

\paragraph{View-dependent effects}
Another limitation faced by single view-port based panorama image view synthesis is the limited view-dependent effects. This becomes more visible in highly glossy and reflective scenes. \Cref{fig:vd} provides an example that visually compares \textbf{OURS}' advancement in producing realistic view-dependent effects.

\subsection{Performance}\label{sec:study:intra}
Virtual and augmented reality demands high frame rates along with high quality to ensure an immersive and comfortable experience.
\rev{Our neural synthesis method achieves real-time performance through our egocentric neural representation (\Cref{sec:method:representation}) as well as the spatial and angular (stereoscopic) foveation (\Cref{sec:synthesis}).}
Further, our latency-quality joint optimization for the representation and network also balances quality and performance (\Cref{sec:method:optimization}).
Here, we evaluate the performance of each component and compare with existing neural synthesis solutions.

For high resolution ($1440\times1600$), high field-of-view ($110$ deg) stereo images required by VR display, our system completes all computation (including gaze-contingent neural-inference and elemental images composition) in $31.8$ms per frame without stereo foveation, as shown in \Cref{tbl:ablation}. \nothing{Compared with {\bf NeRF} and {\bf PANO} (the bottom rows of \Cref{tbl:ablation}), our spatial foveation significantly improves the system performance from offline computation (in seconds) to interactive speed (about 30FPS).}\rev{While our egocentric neural representation (which costs $562$ms) significantly speedup the synthesis compared to {\bf NeRF} and {\bf PANO} (the bottom rows of \Cref{tbl:ablation}), our spatial foveation further improves the system performance from offline computation to interactive speed (about 30FPS).} 
\rev{Finally, our full method with both spatial and stereoscopic foveation}\nothing{Our stereoscopic foveation further reduces the computation time and} achieves a high performance \rev{of}\nothing{to} 50 FPS, contributing to a temporally continuous viewing experience without loss of perceived resolution and quality.

\begin{table}[thb]
    \centering
    \begin{tabular}{wl{5cm}|wr{2cm}}
        \toprule
        foveal infer (per eye)  & 8 \\
        \midrule
        mid- \& far-periphery infer (per eye) & 5.4 \\
        \midrule
        blending \& contrast enhancement & 0.1 \\
        \midrule\midrule
        \rev{{\bf OUR} method without foveation} & \rev{562} \\
        {\bf OUR} method without stereo foveation & 27 \\
        {\bf OUR} full method & {\bf 21.5} \\
        \midrule
        {\bf NeRF} \cite{mildenhall2020nerf} & \new{$9.0\times10^4$} \\
        \midrule
        {\bf PANO} \cite{Lin:DeepPanorama} & \new{$1.0\times10^4$} \\
        \bottomrule
    \end{tabular}
    \Caption{Time consumption breakdown and comparison.}
    {The numbers show the average time consumption (in ms) of each component and the overall system per frame. All units are in millisecond.}
    \label{tbl:ablation}
\end{table}





\section{Conclusion}
\label{sec:conclusion}
    
    
    
\begin{figure}[tb]
    \subfloat[room scene]{\includegraphics[width=0.48\linewidth]{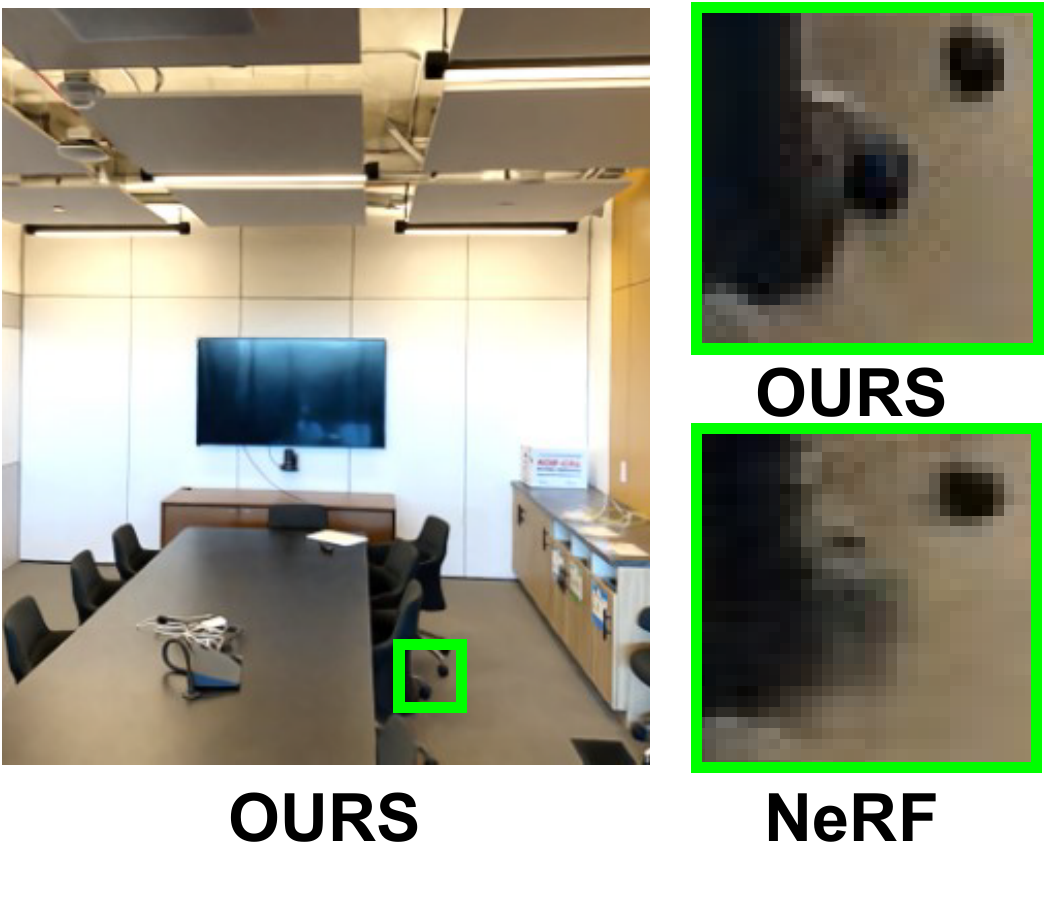}}\label{fig:results:room}
    \subfloat[trex scene]{\includegraphics[width=0.48\linewidth]{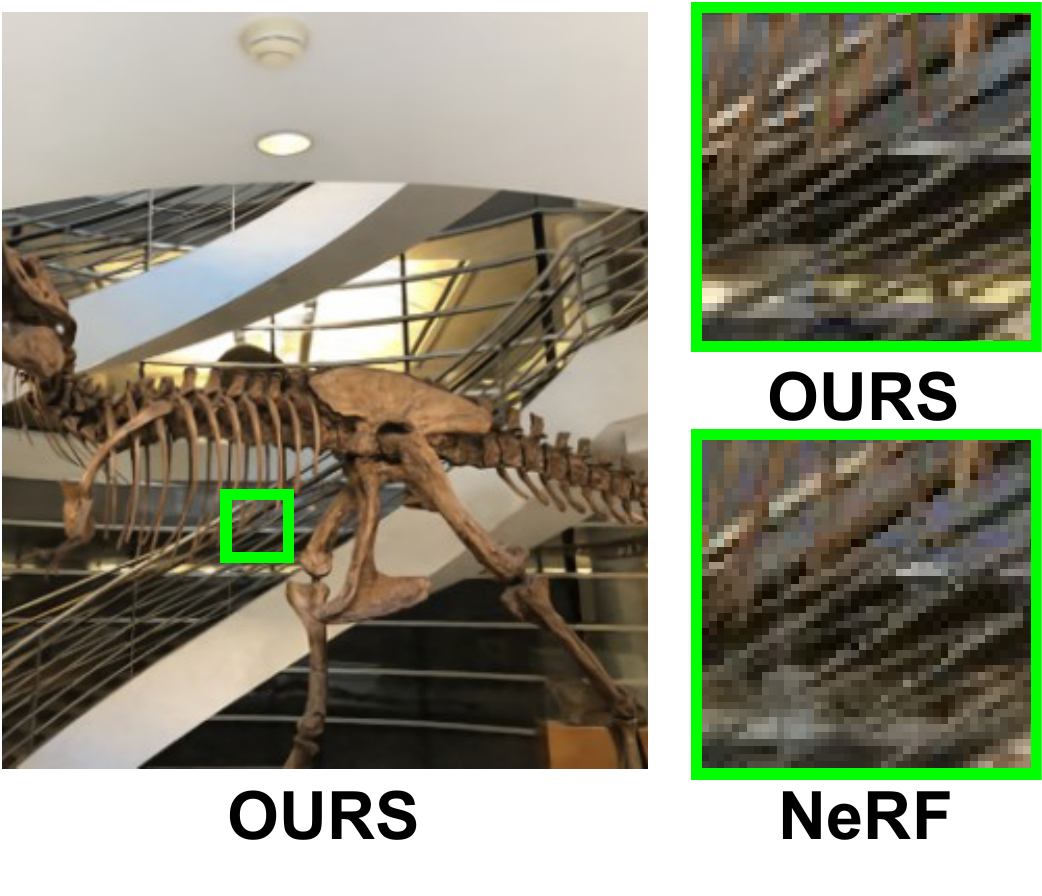}}\label{fig:results:trex}\hspace{-2em}
    \Caption{Neural rendering with physically captured scenes.} 
    {%
    With the data from \protect\cite{mildenhall2020nerf}, we compare the visual quality between {\bf OURS} and {\bf NeRF}. Due to the limited data field-of-view coverage, we trained the model with fovea-only network parameters.
    The zoom-in images further visualize {\bf OURS}' enhanced quality in high depth disparity and high frequency areas.
    \qisun{(May 18) TODO: Include some PSNR overlays on the figures.}
    }
    \label{fig:results:comparison2}
\end{figure}

We present a gaze-contingent neural scene representation and view synthesis method tailored for egocentric and stereoscopic VR viewing. 
To unlock the practical deployment of neural radiance fields in VR, we overcome their challenges such as high latency, low resolution, and low fidelity that are exacerbated with stereoscopic and immersive displays.
This is achieved by encoding not only the scene content but also \textit{how} human vision perceives it, i.e., the gaze-contingent visual acuity and stereopsis.
Our network individually synthesizes foveal, mid-, and far-periphery retinal images, and then blended them to form a wide field-of-view image. 
We also derive an analytical model depicting quality-latency balance and optimizes these two essential factors based on psychophysical study data.
Orthogonal to traditional rendering pipeline, our method takes advantage of NeRF's versatile capability of synthesising not only virtual content but also physically captured scenes, as demonstrated in \Cref{fig:results:comparison2}.
Compared with NeRF \cite{mildenhall2020nerf}\nothing{alternative neural view synthesis approaches (\cite{mildenhall2020nerf})}, our method 
creates significantly faster and higher perceptual fidelity for high resolution and high FoV immersive viewing.
Furthermore, with the support of view-dependent effects (\Cref{fig:vd}), our method is robust to occluded scenes with complex geometry, a challenge faced by panorama-based synthesis approaches (e.g., \cite{Lin:DeepPanorama}).

\paragraph{Limitations and future work}
\qisun{(09/17) NOTE to myself: discuss dynamic scenes.}
While our method achieves superior performance compared to existing approaches, its broader deployment requires combating several constraints.
\new{Our multi-spherical representation renders optimal quality when the virtual camera is within the innermost sphere. 
Although our method unlocks large translation scale than panorama + neural dis-occlusion approaches\qisun{(May 19) Make sure we include example either as a small fig or video}, the optimal translation range is constrained by the radius of this sphere.
However, infinitely enlarge it will decrease synthesized image quality.
}
Therefore, devising an adaptive and dynamic system that automatically optimizes the coordinates may shed lights on allowing traversal-scale translation range without decreasing the perceptual quality or performance.
Similarly, multiscale coordinates that consider various level-of-details of the 3D space have shown their effectiveness of interpolating geometries \cite{winkler2010multi}. 
Developing a corresponding network that synthesizes the imagery from global to local level-of-details would be an interesting direction for future work that extends the applicability to large-scale scenes with strong occlusions.

\rev{Aliasing in the periphery exists because of the low sampling rate. Although we have not apply any anti-aliasing methods yet, the result of our user study shown that the aliasing doesn't cause significant degradation of perceptual quality. While traditional anti-aliasing methods (such as MSAA used in Guenter et al.\cite{Guenter:2012:F3G}) may have an effect, these methods will introduce a significant performance penalty when combined with neural synthesis methods. We think that a candidate solution may be applying multi-scale-encoding, as proposed by Barron et al.\cite{barron2021mipnerf,barron2022mipnerf360}.}

We sample the scene fully based on eccentricity, considering acuity and stereopsis. However, we envision that fine-grained visual sensitivity analysis, such as luminance \cite{Tursun:2019:LCA} or depth \cite{Sun:20:OE}, would provide more insights on achieving higher quality and/or faster performance. 

For simplicity, the spatial-temporal joint optimization in \Cref{sec:method:optimization} connects the output precision and latency to the number of the spheres ($\sphereNum$) but not their radii $\mathbf{\sphereRadius}$. This is due to the potentially significantly higher parameter sampling. Incorporating the parameters into a single training process may significantly reduce the time consumption for the optimization. With the adaptive training process, a content-aware distribution (i.e., $\mathbf{\sphereRadius}$) of the spheres would further improve the synthesis quality and performance. 
\new{The perceptually-based method is orthogonal yet compatible with other remarkable neural radiance field inference acceleration solutions, e.g., \cite{autoint,Wizadwongsa2021NeX}. 
Combining varies perspectives may unlock the future instant and high quality immersive viewing experience such as teleportation.}

\acknowledgments{
This work was partially supported by the National Key Research and Development Program of China (2018YFB1004902).}

\bibliographystyle{abbrv-doi}

\bibliography{paper}

\ifthenelse{\equal{\final}{0}}
{
\clearpage
\pagenumbering{roman}

\input{note}
}
{}

\end{document}